\documentclass[12pt,sort&compress]{elsarticle}
\usepackage[utf8]{inputenc}

\usepackage{bm}
\usepackage{placeins}
\usepackage{booktabs} 
\usepackage{mathrsfs}
\usepackage{amsmath}
\usepackage{amsfonts}
\usepackage{amsthm}
\usepackage{upgreek}
\usepackage{color}
\usepackage{sansmath}
\usepackage{lscape}
\usepackage{adjustbox}

\usepackage{array}

\usepackage{floatrow} 
\usepackage{graphicx}
\usepackage[tableposition=top]{caption}

\usepackage{caption}
\usepackage{subcaption}
\usepackage{float}
\usepackage{url}
\usepackage{multicol}
\usepackage[top=1in,bottom=1in,left=1in,right=1in]{geometry}
\usepackage[pdfborder={0 0 0},colorlinks,allcolors=blue]{hyperref}
\usepackage{txfonts}
\usepackage{setspace}
\usepackage{arydshln}
\usepackage{enumitem}
\usepackage{lipsum}
\usepackage{siunitx,booktabs}
\usepackage{nth}
\usepackage{accents} 
\usepackage{multirow} 
\usepackage{float}  
\usepackage{floatrow} 

\usepackage{algorithm}
\usepackage{algpseudocode}
\usepackage{mdframed}
\usepackage{caption}
\usepackage{xcolor}

\theoremstyle{definition}

\usepackage{tikz}
\usetikzlibrary{shapes.geometric, arrows.meta, positioning}

\tikzstyle{startstop} = [rectangle, rounded corners, minimum width=3.5cm, minimum height=1cm, text centered, draw=black]
\tikzstyle{process} = [rectangle, minimum width=4.5cm, minimum height=1cm, text centered, draw=black]
\tikzstyle{decision} = [diamond, aspect=2, text centered, draw=black]
\tikzstyle{arrow} = [thick,->,>=stealth]
\allowdisplaybreaks

\usepackage{prettyref}
\newrefformat{fig}{Figure~\ref{#1}}
\newrefformat{Fig}{Fig.~\ref{#1}}
\newrefformat{Eq}{Eq.(\ref{#1})}
\newrefformat{eq}{Eq.(\ref{#1})}
\newrefformat{sec}{Section~\ref{#1}}
\newrefformat{tab}{Table~\ref{#1}}
\newrefformat{Tab}{Table~\ref{#1}}
\newrefformat{alg}{Algorithm~\ref{#1}}
\newrefformat{Alg}{Algorithm~\ref{#1}}

\usepackage[dvipsnames]{xcolor}
\newcommand\hl[1]{%
	\bgroup
	\hskip0pt\color{red!80!black}%
	#1%
	\egroup
}

\definecolor{shadecolor}{cmyk}{0,0,0,0.03}

\newmdenv[
  topline=false,
  bottomline=false,
  rightline=false,
  linewidth=2pt,
  leftmargin=0cm,
  linecolor=black,
  backgroundcolor=gray!5,
  skipabove=10pt,
  skipbelow=10pt,
  innertopmargin=5pt,
  innerbottommargin=5pt,
  innerleftmargin=10pt,
  innerrightmargin=10pt
]{algoBox}

\usepackage{amssymb}

\usepackage{tikz}
\usetikzlibrary{shapes.geometric, arrows.meta, positioning}

\tikzstyle{startstop} = [rectangle, rounded corners, minimum width=3cm, minimum height=1cm,text centered, draw=black]
\tikzstyle{process} = [rectangle, minimum width=3cm, minimum height=1cm, text centered, draw=black]
\tikzstyle{decision} = [diamond, aspect=2, text centered, draw=black, inner sep=0pt, minimum width=3cm, minimum height=1.2cm]
\tikzstyle{arrow} = [thick,->,>=stealth]

 \usepackage{lineno}

\def\drawline#1#2{\raise 2.5pt\vbox{\hrule width #1pt height #2pt}}
\def\spacce#1{\hskip #1pt}
\def\solid{\drawline{24}{.8}\nobreak }
\def\bdash{\hbox{\spacce{1}\drawline{4}{.5}\spacce{1}}}
\def\dashed{\bdash\bdash\bdash\bdash\nobreak }

\journal{Applied Energy}

\begin{document}
\begin{frontmatter}




\title{{\normalsize\textit{This article has been submitted to Applied Energy. Final version will appear at} \url{https://www.sciencedirect.com/journal/applied-energy}}\\[0.5em]
Wind farm layout optimization using a novel machine learning approach}

\author[sbu]{Mehrshad Gholami Anjiraki}
\author[sbu]{Christian Santoni}
\author[sbu]{Samin Shapourmiandouab}
\author[sbu]{Hossein Seyedzadeh}
\author[sbu]{Jonathan Craig}
\author[penstate]{Fotis Sotiropoulos}

\author[sbu]{Ali Khosronejad\corref{cor1}}
\cortext[cor1]{Corresponding author: \href{mailto:ali.khosronejad@stonybrook.edu}{ali.khosronejad@stonybrook.edu}}

\address[sbu]{Department of Civil Engineering, Stony Brook University, Stony Brook, NY 11794, USA}
\address[penstate]{Pennsylvania State University, University Park, PA 16802, USA}

\begin{abstract}
We present a novel approach to optimize wind farm layouts for maximum annual energy production (AEP). The optimization effort requires efficient wake models to predict the wake flow and, subsequently, the power generation of wind farms with reasonable accuracy and low computational cost. Wake flow predictions using large-eddy simulation (LES) ensure high fidelity, while reduced-order models, e.g., the Gaussian-curl hybrid (GCH), provide computational efficiency. We integrate LES results and the GCH model to develop a machine learning (ML) framework based on an autoencoder-based convolutional neural network, allowing for a reliable and cost-effective prediction of the wake flow field. We trained the ML model using high-fidelity LES results as the target vector, while low-fidelity data from the GCH model serve as the input vector. The efficiency of the ML model to predict the AEP of the South Fork wind farm, offshore Rhode Island, was illustrated. Then, we integrated the ML model into a greedy optimization algorithm to determine the optimal wind farm layout in terms of turbine positioning. The optimized wind farm layout is shown to achieve a 2. 05\% improvement in AEP over the existing wind farm.

\section*{Highlights}
\begin{itemize}
   \item A novel machine learning algorithm is proposed that allows efficient power prediction of utility-scale wind farms.
   \item Validation of the machine learning model against large-eddy simulation results for an unseen utility-scale wind farm illustrates the reliability of the model.
   \item The machine learning model is integrated into the Greedy algorithm to allow for precise and low-cost optimization of turbine positioning within a wind farm.
   \item The integrated optimization approach obtains the optimal positioning of individual turbines within a farm to maximize its annual energy production.
   \item Applying the optimization approach to an existing offshore wind farm, it was shown that the optimized wind farm layout achieves over 2\% improvement in annual energy production.
\end{itemize}
\end{abstract}

\begin{keyword}
{Wind farm optimization, Greedy algorithm, Large eddy simulation, Wind farm modeling, Machine learning}
\end{keyword}

\end{frontmatter}

\section{Introduction}\label{sec:intro}
\noindent Over the millennium-long history of wind energy \cite{[173],[2]}, considering it as a widespread renewable resource is a relatively recent development. From 1990 to 2021, wind energy generation rose from 3.6 billion kWh in 16 countries to $1808$ billion kWh in 128 countries~\cite{[2]}. \citet{[3]} projected that wind energy will continue to grow from 30\% in 2023 to 46\% of renewable energy production in 2033, surpassing hydropower to become second only to solar energy. To become a competitive global renewable resource, there has been an exponential optimization effort in turbine modeling since the latter half of the 20th century. As asserted by \citet{[65]}, wind farm layout optimization is pivotal to the efficient use of wind power as a global renewable resource. The simple analytical wake model established by~\citet{[4]} served as the foundation for subsequent models aimed at optimizing wind farm power generation and advancing optimization technologies~\cite{[166]}. By the end of the 1990s, \citet{[6]} compiled a survey of wake modeling methods for wind turbine design. According to~\citet{[7]}, individual wind turbines in the 2000s nearly approached the theoretical Betz limit~\cite{[127]}, but there had not been enough research on wind farm optimization to reduce interference between turbines. Various studies reported that unoptimized turbine layout suffered 10-20\% of power losses~\cite{[8],[9]}, 10-40\%~\cite{[7]}, or as much as 60-80\% in extreme cases~\cite{[10]}.

Turbine spacing is not only a fundamental problem but also a multifaceted challenge in offshore wind farm layout optimization. There are natural and practical constraints on wind farm layout, including seabed morphology, national boundaries, and commercial fishing zones, which entail turbine wake interactions with lower velocity and higher turbulence from upstream to downstream~\cite{[11],[12],[13]}. Excessive lateral wake displacements could cause severe power penalties on top of reduced available energy for downstream turbines~\cite{[14]}. However, optimal turbine spacing in an array might significantly raise power production~\cite{[9],[15],[16],[17]}. 

Until the end of the $2000s$, there was a discrepancy between contemporary wind farm design and the development of wind farm optimization theory and measurements, whereby wind farms in practice often had smaller spacing between turbines than what \citet{[68]} found to be optimal. Albeit limited at the time, the growing availability and integration of computational modeling catalyzed layout optimization at the level of multiple turbines~\cite{[37],[63],[64],[65],[66],[68]}. Other studies on layout optimization by computer modeling investigated the intersection between layout optimization and near-wake aerodynamic behavior~\cite{[67]}, modification of wind energy software tools to test standards~\cite{[37],[64]}, and even machine learning for more efficient optimizations~\cite{[65],[66]}. \citet{[69]} extended the analysis of \citet{[68]} to underscore the significance of wind farm length in optimal turbine spacing. For the iterative design process of wind farm layouts in computational models, computational efficiency and accurate rendering of design parameters and control features are required~\cite{[45]}. Hence, the determination of the ideal balance between wind farm layout and turbine wake interaction has remained the primary optimization problem. 

 Analytical turbine wake models are widely used to estimate wake velocity deficits, with upsides such as low computational cost and emphasis on physics through conservation of fundamental flow properties~\cite{[17],[18],[11],[19],[20],[21],[13]}. Until the mid-2000s, computational fluid dynamics (CFD) was not economically viable due to the high computational cost and limited development~\cite{[22]}. By as late as 2010, \citet{[23]} suggested that CFD techniques such as large-eddy simulation (LES) could supplement simple analytical predictions with higher resolution of turbine wakes, which is essential for wake control and wind farm optimization~\cite{[24],[11],[25]}. Such shortcomings provided an opportunity for the development of CFD simulation approaches to achieve high-fidelity wake modeling that is not attainable by analytical methods alone~\cite{[26],[163],[28],[11]}. CFD models elucidate turbine wake dynamics including wake meandering, atmospheric stability, influence from topology, and effects from the tower and nacelle~\cite{[29],[20],[63]}. Moreover, the actuator methods in CFD lower computational demand by simplifying the turbine geometry and the required resolution of the flow field~\cite{[31]}. These methods include the actuator disk (AD), actuator line (AL), and actuator surface (AS). All these methods respectively offer increasing fidelity for more computational demand and detail~\cite{[32],[33],[34],[35],[31]}. Nonetheless, there has been an inherent obstacle due to the greater computational cost of high-fidelity approaches such as LES.

 Recently, the integration of analytical and CFD models has emerged as a central approach in wind farm analysis, aiming to balance computational efficiency, physical fidelity, and detailed flow field prediction. Analytical models offer low computational demand, while CFD methods, particularly LES, provide high-resolution insights into wake dynamics. Their combined use enables researchers to efficiently capture complex wake dynamics without compromising accuracy. For instance, \citet{[29]} combined a turbine model using a 2D elliptical Gaussian distribution with LES to elucidate the influence of the atmospheric thermal stability on the wake of the wind turbines. Through high-fidelity modeling of a small wind farm, \citet{[36]} showed that yaw optimization using a control system raised total energy production. \citet{[28]} found reasonable agreement between their new model, LES, and power measurements for an overall improvement in wind farm power prediction. Later, analytical models coupled with CFD have maintained a strong focus on wind farm optimization. For example, \citet{[13]} reported accurate predictions of wake width and maximum velocity deficit by a physics-based analytical model, upon validation by LES of a utility-scale wind turbine wake.

Coupled with analytical models and CFD simulations, control systems have recently become a recurring theme in wind farm optimization. Control systems often apply wake steering for yaw misalignment in upstream turbines to maximize the turbine array energy production~\cite{[12],[46]}. The work of \citet{[23]} set the stage for future wake steering in analytical modeling of wind farm optimization, where they investigated the viability of wake steering through LES. Many studies improved wind farm control by adapting various models found in FLORIS (FLOw Redirection and Induction in Steady State) as an analytical model ~\cite{[36],[37],[49],[50],[51],[43]}. With a focus on yaw as an optimization parameter, \citet{[36]} developed a parametric model for wind plant control optimization in an online implementation with coupled LES and actuator line. \citet{[37]} and \citet{[49]} also used FLORIS with the Jensen wake model \cite{[4]} to increase the power density of wind farms using a yaw-control approach. Furthermore, \citet{[50]} demonstrated through LES that engineering models had not captured the influence of the counter-rotating vortices on wake steering. \citet{[43]} used wind direction measurements to reduce controller delay in wind farm wake steering. Building on the Gaussian model elaborated by \citet{[26], [163]}, \citet{[51]} extended the curled wake mechanism to synthesize a control-oriented curled wake model known as the Gaussian curl hybrid (GCH) model. \citet{[53]} displayed the viability of the GCH model by achieving higher agreement with LES than what the standard Gaussian method achieved. Even after coming so far, the GCH model suffers shortcomings such as struggling to accurately capture the momentum deficit and wake recovery~\cite{[38],[11],[44],[25]}. Studies such as those of \citet{[45]} and \citet{[25]} addressed the limitations of the GCH by developing a cumulative curl (CC) model to include more wake predictions. Overall, as \citet{[44]} assessed, model validation remains a substantial obstacle to commercial wind farm control testing. The active development of wind farm optimization methods and technology indicates more potential for refinement and incorporation in state-of-the-art practice.

Coinciding with advancements in wind energy modeling, foundational developments in both machine learning (ML) and optimization methods began to emerge. These parallel improvements laid the groundwork for recent progress in velocity field prediction and wind farm layout optimization.
For example, \citet{[55]} produced the ADAM method to solve the objective function by first-order, gradient-based optimization. About the same time, \citet{[54]} verified a specialized convolutional neural network (CNN) architecture known as U-Net, which outperformed the sliding-window convolutional network in computational speed and quality of the outputs. Due to their broad applications and utility in pre-existing solutions, ML techniques have supplemented coupling of analytical models and CFD-based simulations. In a comparison of ML techniques for wind farm analysis, \citet{[56]} found that more elaborate models, e.g.,  artificial neural network (ANN) and support vector regression (SVR) models, could predict wind farm power production more accurately.

By the 2020s, \citet{[60]} and \citet{[61]} demonstrated that the ANN and deep neural network (DNN) accurately and efficiently predicted wind farm behavior from AD and experimental data, respectively. Whereas \citet{[47]} warned that LES might be too computationally costly for optimization of yaw control and wake interactions in turbine arrays. The autoencoder convolutional neural network (ACNN) of \citet{[25]} reduced prediction error from $20\%$ in the GCH model to less than 5\%, with comparable cost to an analytical model and accuracy to high-fidelity LES. \citet{[62]} reported that a CNN trained on high-fidelity simulation data successfully predicted both the time-averaged flow field around individual turbines and their power output, achieving nearly $90\%$ reduction in computational time.
While ML techniques have primarily been used for flow field and power output prediction, separate optimization strategies, both traditional and learning-based, have been applied to enhance wind farm performance. For instance, \citet{[66]} utilized a greedy algorithm to optimize wind turbine positioning in a two-dimensional domain, incorporating various modeling components such as linear wake effects, Weibull-distributed wind conditions, and profit functions. \citet{[67]} and \citet{[65]} applied the greedy algorithm to various wake models.
Moreover, \citet{[58]} and \citet{[59]} employed reinforcement learning (RL) techniques for one turbine and a multiplier algorithm together with RL to predict wind conditions from optimized yaw angles.

While AI has been successfully applied in previous studies to predict the results of high-fidelity CFD simulations, challenges remain in improving its accuracy for complex systems such as large-scale wind farms, as well as ensuring scalability and seamless integration into a practical optimization framework. Simplified wake models lack sufficient accuracy for layout optimization, whereas high-fidelity simulations, including LES, are prohibitively computationally demanding for iterative optimization tasks. Therefore, a balance can be achieved by coupling low- and high-fidelity data through ML techniques. 

To address this challenge, the present study integrates the low-fidelity velocity fields obtained from the GCH wake model, implemented in FLORIS v3.4, with high-fidelity LES data using an ACNN-based model to optimize the power production of a real-life wind farm, namely the South Fork wind farm, located offshore the coast of Rhode Island. The ML model is trained using the high-fidelity LES data, where the input and the target vectors are obtained from the GCH wake model and LES results, respectively. The trained ML model is then utilized in conjunction with the cost-effective, low-fidelity GCH wake model for inference. Integrating the ML with the greedy algorithm allowed us to optimize the wind farm's layout. We note that the high-fidelity numerical simulations were conducted using the LES module of an in-house model, the Virtual Flow Simulator (VFS-Geophysics) code \cite{[72], [147]}. The AL and AS methods were employed to represent the turbine blades and the nacelle, respectively.

Initially, the performance of the GCH model was evaluated against the LES results, considered as the ground truth, across $14$ distinct wind farm layouts, which are considered as testbeds and obtained by imposing the incoming wind toward the wind farm at different angles. These testbeds served to train the ML model, which learned to map the GCH-derived low-fidelity flow fields to their high-fidelity LES counterparts. The model's generalization capability was tested on an unseen wind farm layout, which was an extension of a study by \citet{[25]}. After validating the predictions against the LES results, the trained ML model was integrated into a greedy optimization framework and applied to the South Fork wind farm. 

Beginning with a randomly initialized turbine layout, the greedy algorithm iteratively updates turbine positions on a predefined grid under various locally measured wind directions and speeds. At each iteration, the ML-predicted high-fidelity velocity fields were used to compute the annual energy production (AEP). The turbine configuration that yielded the highest energy output was retained. The resulting optimized layout was then compared to the baseline configuration in terms of turbine placement and AEP. To further assess the reliability of the ML-based predictions, the optimized layout was also evaluated against the LES simulations under various wind conditions. While previous studies have applied greedy algorithms to wind farm optimization, they often relied on simplified two-dimensional domains \cite{[66]} or low-fidelity wake models \cite{[65]}. To our knowledge, this study is the first to incorporate a greedy optimization algorithm with an ML model trained on LES data, enabling more accurate and computationally feasible wind farm layout design.

The remainder of this paper is organized as follows. \prettyref{sec:2} provides the governing equations of the wake flow field.  \prettyref{sec:2222} describes the machine learning algorithm and its training procedure. In \prettyref{sec:3}, we discuss the details of the test cases, followed by a description of the computational setup in \prettyref{sec:38888888888}. The simulation results are presented and discussed in \prettyref{sec:4}. Lastly, we conclude the findings of the study in \prettyref{sec:5}.

\section{Governing equations} \label{sec:2}
\subsection{Aerodynamics}
\noindent The aerodynamics model of the VFS-Geophysics code solves the spatially filtered Navier–Stokes equations for incompressible flow within non-orthogonal generalized curvilinear coordinates. With compact tensor notation where repeated indices indicate summation, the governing equations are expressed as follows~\cite{[71], [72], [135], [147], [148]}:

\begin{eqnarray}
J\frac{\partial U^j}{\partial\xi^j} &=& 0, \label{eq:1} \\
\frac{1}{J} \frac{\partial U^i}{\partial t} &=& \frac{\xi_l^i}{J} \left[ \frac{\partial}{\partial\xi^j} \left(U^ju_l\right) + \frac{1}{\rho} \frac{\partial}{\partial\xi^j} \left( \mu \frac{g^{jk}}{J} \frac{\partial u_l}{\partial\xi^k} \right) - \frac{1}{\rho} \frac{\partial}{\partial\xi^j} \left( \frac{\xi_l^jp}{J} \right)-\frac{1}{\rho} \frac{ \partial \tau_{lj}}{ \partial \xi^j} + F_{\mathrm{ext}} \right], \label{eq:2}
\end{eqnarray}
where the Jacobian of geometric transformation, $J=\left|\partial\left(\xi^1,\xi^2,\xi^3\right)/\partial\left(x_1,x_2,x_3\right)\right|$, handles the transformation from Cartesian to  curvilinear coordinates. $U^i=(\xi_m^i/J)\ u_m,$ is the contravariant volume flux, in which $\xi_l^i=\partial\xi^i/\partial x_l $. $u_l$ is the $l$-th filtered velocity component in Cartesian coordinates. $\mu$ is the dynamic viscosity of air, $\rho$ is the density of air, and $G^{jk}=\xi_l^j\ \xi_l^k$ represents the contravariant metric tensor,  $p$ denotes the pressure, $F_{\mathrm{ext}}$ includes the lift and drag as forces per unit volume. The dynamic Smagorinsky model captures the subgrid-scale stresses in the LES turbulence model~\cite{[100], [103], [104]}, which is defined as such: 
\begin{eqnarray}
     \tau_{lj} &=& -2\mu_{\mathrm{t}} \overline{S}_{lj}+\frac{1}{3}\tau_{kk}\delta_{lj}, \label{eq:3} \\
    \mu_{\mathrm{t}} &=& C_{\mathrm{s}}\Delta^2 \left |\overline{S} \right|, \label{eq:4}\\
     \left |\overline{S} \right| &=& \sqrt{2 (\overline{S_{lj}} \overline{S_{lj}} )}, \label{eq:5}
\end{eqnarray}
where $\mu_{\mathrm{t}}$ is the eddy viscosity, $\overline{S}_{ij}$ is the filtered strain-rate tensor, \( \delta_{ij} \) is the Kronecker delta, and $C_\mathrm{s}$ is the Smagorinsky coeffificent. The filter size is defined as $\Delta = J^{-1/3}$~\cite{[100]}.

\subsection{Turbine modeling }
\noindent Two widely adopted strategies are commonly used to simulate flow–turbine interactions~\cite{[143]}.  The first strategy entails turbine parameterization techniques: namely, the actuator models introduce lift and drag as body forces into the flow governing equations to represent the turbine's influence as momentum sinks  \cite{[33], [137], [138]}. The second strategy is the geometry-resolving approach \cite{[139], [140], [142], [178]}, which operates by resolving the turbine blades' geometry with fine computational grids to simulate fluid-structure interaction. Although the latter approach delivers higher fidelity with minimal simplifications \cite{[143]}, the heightened computational demand discourages its deployment for turbine arrays. Hence, we employ the AL method, which balances the computational efficiency and degree of detail in turbine-induced flow features.

The AL method interprets each wind turbine blade as a rotating straight line that projects body forces onto the flow field. Each blade is discretized into radial segments, and the lift and drag forces on each segment are computed as follows~\cite{[137]}:
\begin{eqnarray}
  F_{\mathrm{L}} &=& \frac{c}{2}\rho C_{\mathrm{L}}V_{\mathrm{rel}}^2, \label{eq:7}\\
  F_{\mathrm{D}} &=& \frac{c}{2}\rho C_{\mathrm{D}}V_{\mathrm{rel}}^2, \label{eq:6} 
\end{eqnarray}
where $c$ is the turbine blade's chord length, $C_{\mathrm{D}}$ and $C_{\mathrm{L}}$ denote the drag and lift coefficients, respectively. The relative velocity, $V_{\mathrm{rel}}$, is obtained as follows~\cite{[137]}:
\begin{equation}
V_{\mathrm{rel}} = \left(u_x, u_\theta - \Omega r\right),
\label{eq:8}
\end{equation}
where $\Omega$ is the angular velocity of the rotor, and $r$ is the radial distance along each AL segment from the rotor center. The velocity components $u_x$ and $u_\theta$ indicate the axial and azimuthal components, respectively, averaged over the actuator line. 

\subsection{Control module}
\noindent The model employs the baseline collective pitch control (CPC) strategy, as described in~\cite{[176]}. In this approach, the rotor's angular velocity is regulated by setting the generator torque to maximize aerodynamic efficiency and minimize the risk of structural damage, as follows:
\begin{equation}
\tau_{g} = \frac{\pi \rho R^5 C_{p,\text{max}}}{2 \lambda_{\text{opt}}^3 G^3}\omega_g^2, \label{eq:16_1}
\end{equation}
where $\tau_{g}$ represents the generator torque, and $R$ is the rotor radius. The term $C_{p,\text{max}}$ denotes the maximum power coefficient of the turbine, while $\omega_g$ is the angular velocity of the generator. $\lambda_{\text{opt}}$ refers to the optimal tip speed ratio, and $G$ is the gearbox ratio. More information can be found in~\cite{[176]}.

\subsection{Analytical wake model}
\noindent Implemented in FLORIS (v3.4), the GCH model~\cite{[165]} provides the input vector to our ML model. The GCH model comprises the Gaussian wake formulation developed by \citet{[26], [163]} and \citet{[162]} as well as the curl-based model proposed by \citet{[51]}. The Gaussian wake model supposes a Gaussian profile for the wake deficit, expressed as:
\begin{equation}
    \frac{u_g}{U_\infty} = 1 - C e^{-[(y - \delta)^2/\sigma_y^2 + z^2/\sigma_z^2]},
    \label{eq:gauss}
\end{equation}
where \( C \) is the centerline velocity deficit coefficient, \( \delta \) is the lateral wake deflection, and \( \sigma_z \) and \( \sigma_y \) are the standard deviations corresponding to the wake width in the wall-normal and spanwise directions. The wake width is determined as follows:
\begin{equation}
    \sigma_i = k_i \Delta x + \epsilon,
    \label{eq:sigma}
\end{equation}
where \( \epsilon \) is the initial wake width at the turbine, \( \Delta x \) is the distance downstream from the rotor, and \( k_i \) is the wake expansion coefficient in the \( i \)-direction. This coefficient reflects the influence of turbulence-driven mixing and atmospheric stability on the growth of the wake. 

The GCH model parameters used herein were adopted directly from the sample input configuration distributed with the \textsc{FLORIS} (v3.4) software package. Albeit not necessarily optimal for all use cases, the study kept the default settings to evaluate the ML model performance without further calibration. More details can be found in \cite{[25]}.

\section{Autoencoder convolutional neural network} \label{sec:2222}
\begin{figure} [H]
  \includegraphics[width=1\textwidth]{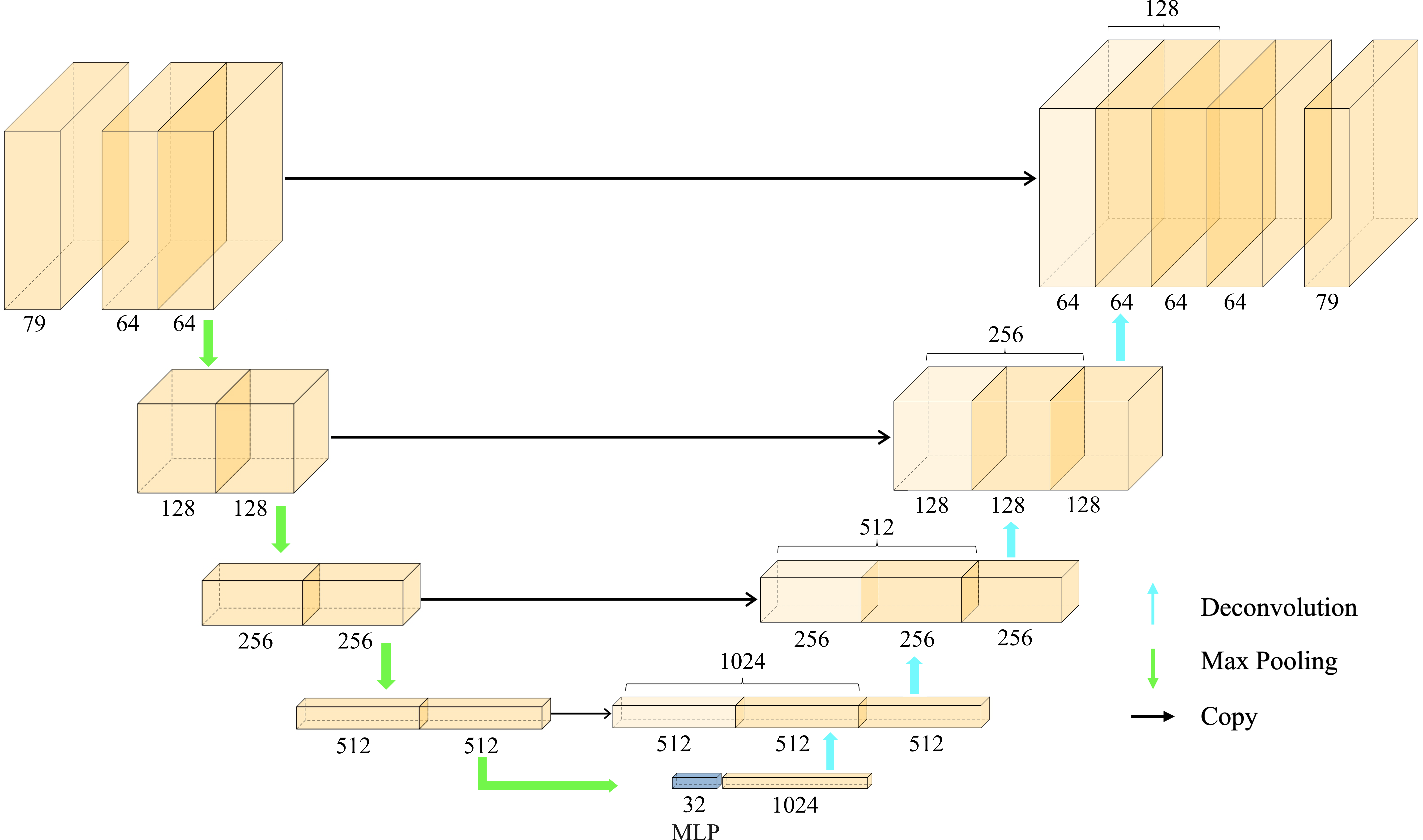}
  \caption{Architecture of the autoencoder-based convolutional neural network combined with a fully connected MLP parameter branch. Arrows indicate the sub-layer operations, including max pooling, deconvolution, and skip connections. The numbers adjacent to each block represent the number of channels (for convolutional layers) or neurons (for fully connected layers) at each stage, including the latent space representation.}
  \label{Fig:1_0}
\end{figure}

\noindent We implemented a convolutional neural network based on the U-Net architecture~\cite{[54]} and augmented it with a lightweight parameter-processing module that combines a multilayer perceptron (MLP) and a transformer encoder~\cite{[177]}. The CNN model takes as input a low-fidelity wake slice generated by the GCH model, with dimensions \( X \in \mathbb{R}^{79 \times L \times W} \), alongside a control vector \( c = [x_t,\; y_t,\; \gamma,\; D,\; U_\infty] \in \mathbb{R}^5 \). Here, \( x_t \) and \( y_t \) represent the turbine’s position in the horizontal plane, \( \gamma \) denotes the yaw angle, \( D \) is the rotor diameter, and \( U_\infty \) is the bulk wind velocity. The output vector of the trained U-Net produces a time-averaged flow field, approximating LES fidelity, with size \( \hat{Y} \in \mathbb{R}^{79 \times L \times W} \). A schematic of the ML model's architecture is plotted in \prettyref{Fig:1_0}.

The proposed ML model is composed of three main components: the parameter branch, the U-Net encoder–decoder core, and the output projection layer~\cite{[25]}. The control vector, \(c\), is first processed through three fully connected layers activated by ReLU functions~\cite{[175]}. The resulting vector is passed through a stack of four transformer encoder layers, yielding a 64-dimensional embedding. This embedding is reshaped and broadcast across spatial dimensions to produce a parameter tensor, which is fed into the bottleneck of the U-Net to inform the spatial encoder features.
The encoder begins with an ``InConv'' block consisting of two \(3 \times 3\) convolutions with ReLU activation, maintaining spatial resolution. This is followed by four downsampling stages, each composed of max pooling and two \(3 \times 3\) convolutional layers with ReLU and dropout (rate = 0.1), progressively increasing the number of feature channels.

At the network's bottleneck, a \(1 \times 1\) convolution followed by ReLU activation is applied to the encoded feature tensor. The result is then combined with the parameter embedding via residual addition, which helps maintain stable gradient propagation (i.e., backpropagation) and enables the network to refine the features more effectively by learning incremental corrections. The decoder mirrors the encoder with four upsampling stages. Each stage performs a transposed convolution to restore spatial resolution, concatenates the output with the corresponding encoder feature map via skip connections, and applies two \(3 \times 3\) convolutions with ReLU and dropout. The final decoder output, with $64$ channels, is passed through a \(1 \times 1\) convolution to produce a $79$-channel prediction representing the LES-equivalent wake field, \(\hat{Y}\). The blockwise configuration of the U-Net model is presented in \prettyref{appendix:app1}.

This architecture leverages the benefits of both visual feature extraction and parameter conditioning. The integration of the parameter embedding via residual projection at the bottleneck promotes stable training and enables the network to make localized corrections without relearning the entire feature representation. Additionally, the symmetric encoder–decoder design and skip connections allow the model to retain fine spatial details while learning multi-scale representations of the flow field. For more details, readers are referred to \cite{[25]}.

The U-Net was trained on a set of $14$ paired GCH and LES (time-averaged) flow fields for different scenarios, and evaluated on a single held-out test case (see, \prettyref{sec:3} for more details). The training was conducted for nearly $60\,000$ epochs using the Adam optimizer with an initial learning rate of \(10^{-4}\) and no weight decay. Each epoch processed the full set of 14 training scenarios. A dropout rate of $0.1$ was applied after each convolutional block to further prevent overfitting. Different parameters of the network were optimized by minimizing the root mean squared error (RMSE) between the predicted (time-averaged) flow field slice \(\hat{Y}\) and the corresponding high-fidelity LES slice \(Y\). Moreover, in order to improve the learning process of the network, we adaptively reduced the learning rate. To do so, we utilized a \textit{ReduceLROnPlateau} scheduler with a reduction factor of $0.5$ and a patience of $100$ epochs and triggered the scheduler based on the current validation RMSE. More specifically, after every $100$ epochs, the model was evaluated on the held-out test scenario by comparing the prediction \(\hat{Y}\) to the ground-truth LES field \(Y\). In this process, if the RMSE of the validation decreased compared to the previous evaluation, the state of the model was checked. Otherwise, the scheduler reduced the learning rate by a factor of 0.5. This adaptive learning rate strategy helps stabilize training, mitigate overfitting, and encourage more effective learning. Lastly, we should note that, in the inference stage, the trained model is evaluated on an unseen test case by forwarding the low-fidelity flow field slice \(X\) as the input vector, which is obtained from the GCH model along with its associated control vector \(c\). The output vector of the trained ML model is then obtained as $\hat Y \;=\; \textit{model}(X, c)$. A summary of the training parameters is provided in \prettyref{tab:0_0}.

\begin{table}
\caption{Training parameters and hyperparameters used during the learning process of the ML model.}
\label{tab:0_0}
\begin{tabular}{@{}p{5.5cm}p{7.8cm}@{}}
\textit{Parameter}         & \textit{Value} \\
\midrule
Training set size          & 14 GHC \& LES datasets\\
Test set size              & 1 held-out dataset \\
Batch size                   &  1 \\
Optimizer                  & Adam, $\mathrm{lr} = 1 \times 10^{-4}$ \\
Scheduler                  & ReduceLROnPlateau \\
Dropout                    & 0.1 \\
Epochs                     & 60,000 \\
Checkpointing              & Checkpoint on RMSE improvement \\
\end{tabular}
\end{table}

\section{Test case description} \label{sec:3}
\begin{figure} [H]
  \includegraphics[width=1\textwidth]{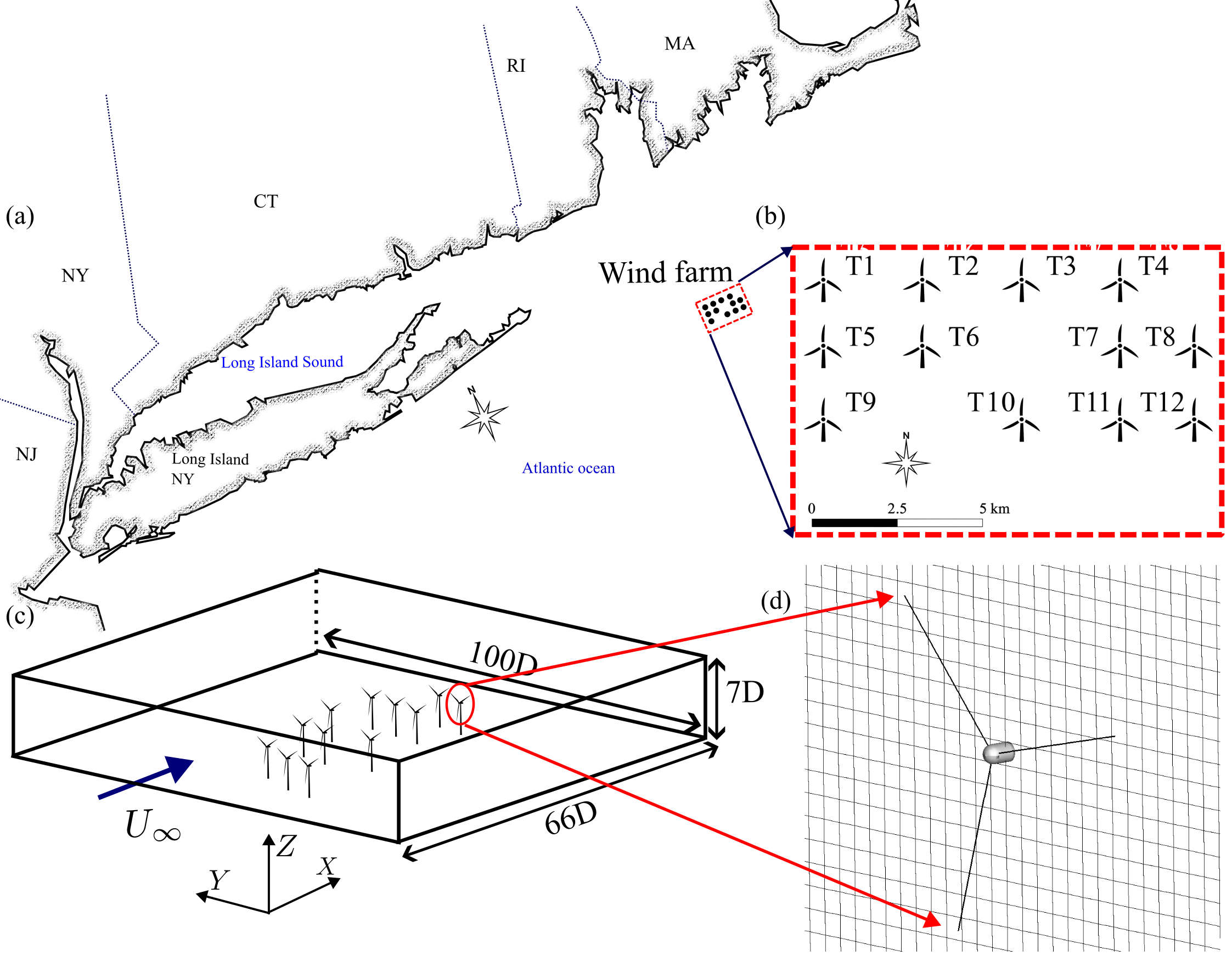}
  \caption{Schematics of the South Fork wind farm geographic location and the computational domain of this study. (a) shows the wind farm's geographic location off the coast of Rhode Island. (b) marks the turbine layout and individual turbine identifiers (T1–T12). (c) depicts the computational domain, which is sized at 100D × 66D × 7D (where D is the rotor diameter), and the inflow direction. The turbine array is arranged to replicate the staggered placement observed in the project site. (d) presents a close-up view of a single turbine embedded within the computational grid, used to resolve near-wake dynamics.}
  \label{Fig:1}
\end{figure}

\noindent This study is focused on the South Fork wind farm as a baseline \cite{[154]}. This wind farm is an offshore wind project located approximately 48 km east of Montauk Point on the South Fork of New York's Long Island. The wind farm comprises $12$ Siemens Gamesa turbines. The turbines are installed in a grid-like configuration aligned with the northing and easting directions, and are spaced approximately $1.5$ km apart. \prettyref{Fig:1} displays the wind farm layout at the actual installation site. This configuration is hereafter defined as the baseline layout. Due to the lack of publicly available aerodynamic specifications for the Siemens Gamesa $11$ MW turbines, the Siemens Gamesa SWT-3.6-120 model was selected for its thorough documentation and usage in high-fidelity wind energy research. The SWT-3.6-120 turbine features a rotor diameter ($D$) of 120 m, a hub height of $0.75D$~\cite{[159]}, a rated wind speed of $12$ m/s, and a rated power output of $3.6$ MW. Furthermore, the turbine operates at an optimum tip speed ratio (TSR) of $\lambda = \Omega D / 2U_{\infty}=9$, where $\Omega$ is the angular velocity and $U_{\infty}$ is the freestream wind speed. The details of the turbines are presented in \prettyref{tab:0}.

\begin{table}[H]
\centering
\begin{tabular}{p{6cm} r}
\textit{Parameter}         & \textit{Value} \\
\midrule
$D$ (m) & 120 \\
Hub height & $0.75D$ \\
Blade length & $0.4875D$ \\
Number of blades & 3 \\
Rated wind speed ($m/s$)& $12$ \\
Rated power output ($MW$)& $3.6$ \\
TSR & 9 \\
\end{tabular}
\caption{Turbine characteristics used for the baseline wind farm. }
\label{tab:0}
\end{table}

\begin{figure} [H]
  \includegraphics[width=1\textwidth]{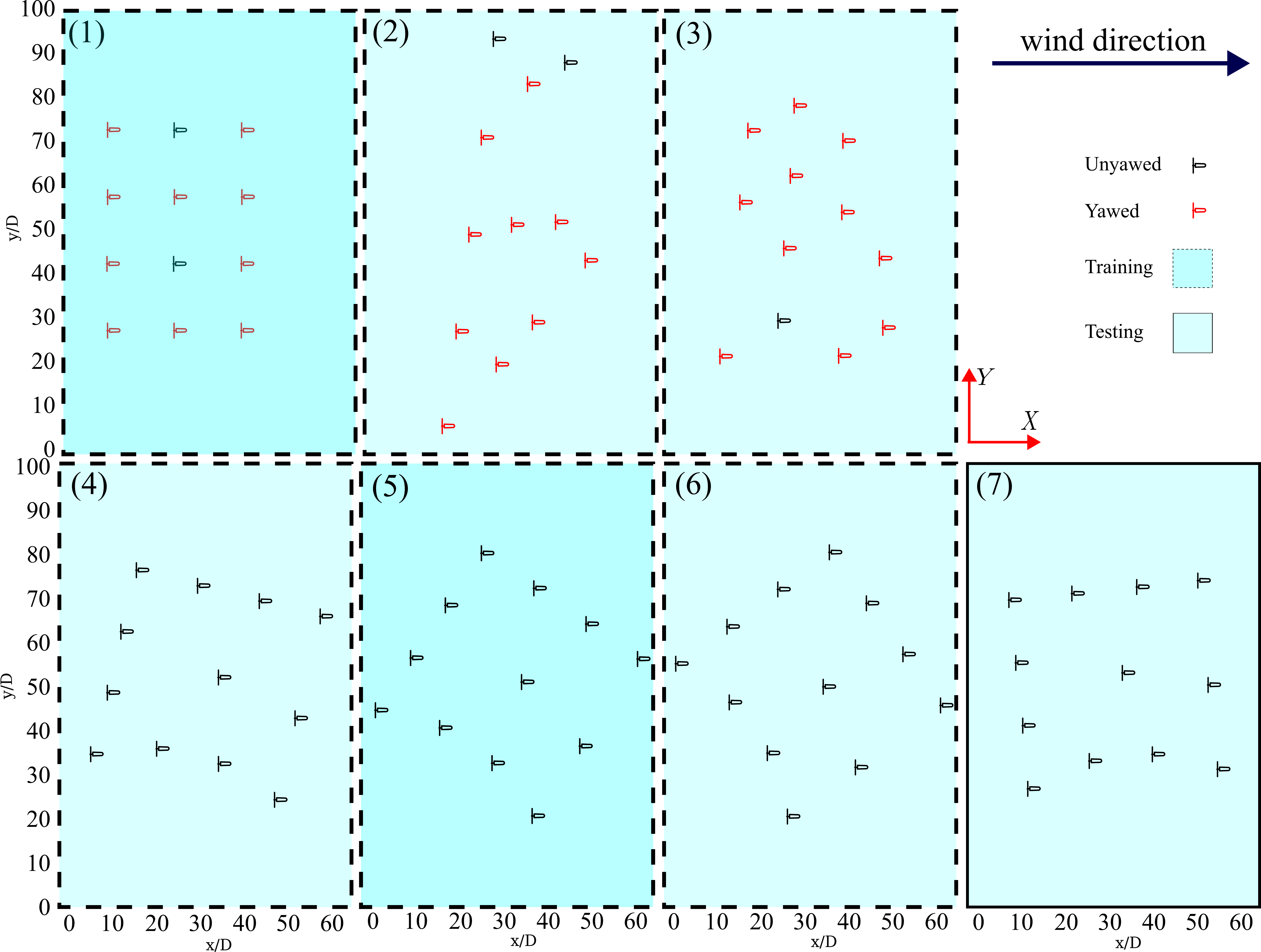}
  \caption{Turbine configurations viewed from the top, used for training (layouts 1 to 6) and testing (layout 7) the ML model. The wind direction is from left to right, and the coordinates are normalized by the rotor diameter ($D$). Black and red symbols mark the unyawed and yawed turbines in each layout.}
  \label{Fig:3_1}
\end{figure}

We considered six distinct turbine layouts (\prettyref{Fig:3_1}(1)-(6)) combined with ten wind speeds, ranging from $5$ to $17 \ m/s$, under both yawed and unyawed conditions. For clarity, the term ``yawed'' denotes wind farm configurations in which both yawed and unyawed turbines could be present. Even though individual yawed turbines may operate at different yaw angles, the specific yaw angles are not detailed and are collectively referred to as ``yawed'' for brevity. It is also important to note that the training dataset could not include all combinations of turbine layouts, wind speeds, and yaw conditions due to computational constraints. As a result, the training dataset comprises $15$ distinct test cases, as described in \prettyref{tab:0_2}. 

The performance of the trained ML model was evaluated on an unseen configurations (\prettyref{Fig:3_1}(7)), featuring a different layout from those used during the training (\prettyref{tab:0_2}). The evaluation aims to assess the generalization capability and robustness of the ML model to predict the velocity fields of wind farms beyond that of the training test cases. 

\begin{table}[H]
\centering
\caption{Test case descriptions. Test cases 1 to 14 are used during the training of the ML model, whereas test
case 15 is designated for testing the performance of the ML model. The turbine configurations are plotted in \prettyref{Fig:3_1}}
\vspace{0.5em}
\begin{tabular}{c c >{\centering\arraybackslash}p{3.1cm} >{\centering\arraybackslash}p{3.2cm}}
\textit{Test case} & \textit{Configuration} & \textit{Wind speed ($m/s$)} & \textit{Yaw} \\
\hline 
1   & \multirow{5}{*}{\centering 1} & 5  & yawed    \\
2   &                              & 8  & yawed    \\
3   &                              & 9  & unyawed  \\
4   &                              & 13 & yawed    \\
5   &                              & 15 & unyawed  \\
\addlinespace
6   & \multirow{3}{*}{\centering 2} & 7  & yawed    \\
7   &                              & 10 & yawed    \\
8   &                              & 13 & unyawed  \\
\addlinespace
9   & \multirow{3}{*}{\centering 3} & 7  & yawed    \\
10  &                              & 11 & yawed    \\
11  &                              & 11 & unyawed  \\
\addlinespace
12  & 4                            & 10 & unyawed  \\
13  & 5                            & 14 & unyawed  \\
14  & 6                            & 17 & unyawed  \\
15  & 7                            & 7  & unyawed  \\
\end{tabular}
\label{tab:0_2}
\end{table}

\section{Computational details} \label{sec:38888888888}
\noindent The computational domain spans $66D$ in the streamwise ($x$), $100D$ in the spanwise ($y$), and $7D$ in the vertical ($z$) direction.
The domain was discretized by $476$, $1679$, and $79$ computational grid nodes along the $x$, $y$, and $z$ directions, respectively, yielding grid resolutions of $0.14D$ (streamwise), $0.06D$ (spanwise), and $0.06D$ (vertical).
The temporal step of the computations was selected such that the Courant-Friedrichs-Lewy stays smaller than one at all times (see \prettyref{tab:4}). We should mention that, in a series of studies, we have extensively validated the flow solver, the actuator line model, the ML model, and documented the sensitivity of the computed flow field to the grid resolution (see, e.g.,~\cite{[62], [25], [71], [72], [34]}).

A periodic boundary condition is applied in the spanwise direction of the flow domain. The Neumann boundary condition is imposed at the outlet. The inflow boundary condition is obtained through a separate precursor simulation in the flow domain without turbine. Considering a periodic boundary condition in the streamwise direction, the precursor simulation generated an instantaneously converged turbulent flow representative of a neutral atmospheric boundary layer. The plateau of total kinetic energy signaled the onset of fully developed turbulence. The instantaneous turbulent velocity field at a cross-plane located at the mid-length of the precursor domain was then extracted and applied as the inlet boundary condition for the wand farm.

The free-slip condition is applied at the top boundary of the flow domain. At the bottom boundary, a logarithmic law of the wall is implemented according to the given definition~\cite{[25]}:
\begin{equation}
U = \frac{U_*}{\kappa} \ln \frac{z}{z_0},
\label{eq:19}
\end{equation}
where \( \kappa \) is the von Kármán constant (\( \kappa = 0.41 \)), \( U_* \) is the shear velocity, and \( z_0 \) denotes the surface roughness height~\cite{[25], [62]}.

Finally, it is important to mention that the computations were conducted on a Linux cluster consisting of several thousands of Intel Skylake CPU cores. On average, LES of each case required $34\,500$ CPU hours to achieve statistically converged flow fields. In contrast, the flow field predictions using the GCH and ML models were performed on a single Intel Skylake CPU core. The GCH and ML models needed only about $0.033$ and $0.036$ CPU hours per test case, respectively. This comparison clearly shows the superior cost effectiveness of the ML model relative to the LES. 

\begin{table}[H]
\centering
\renewcommand{\arraystretch}{1.25}
\setlength{\tabcolsep}{6pt} 
\begin{tabular}{p{6cm} r}
\textit{Parameter}         & \textit{Value} \\
\midrule
$N_x, N_y, N_z$ & $476 \times 1679 \times 79$ \\
$\frac{\Delta x}{D}, \frac{\Delta y}{D}, \frac{\Delta z}{D}$ & $0.14$, $0.06$, $0.06$ \\
$z^+$ & 1320 \\
$\Delta t$ & 0.005 \\
\end{tabular}
\caption{Computational details of the flow solver, which employs a structured computational grid with \( N_x \), \( N_y \), and \( N_z \) grid cells in the streamwise, spanwise, and vertical directions, respectively. Normalized by rotor diameter, \( D \), the spatial resolutions are given as \( \Delta x / D \), \( \Delta y / D \), and \( \Delta z / D \). The vertical grid spacing near the wall is represented in wall units as \( z^+ \). \( \Delta t^* = \Delta t U_{\infty} / D \) is the dimensionless time step, where \( \Delta t \) is the physical time step.}
\label{tab:4}
\end{table}

\section{Findings and discussion}\label{sec:4}
\noindent We begin with a comparative analysis of wake flow prediction results using LES ane GCH model. This analysis is carried out for various turbine configurations, yaw angles, and wind speeds to assess the prediction accuracy of the GCH model. Next, we evaluate the robustness of the ML model predictions by analyzing their consistency and performance relative to LES method on an unseen wind farm layout. Finally, we apply the developed framework to a wind farm optimization case study, demonstrating the practical impact of the proposed modeling and prediction strategies.

\subsection{Evaluating the performance of the Gauss-curl hybrid model}
\begin{figure}[H]
  \includegraphics[width=0.95\textwidth]{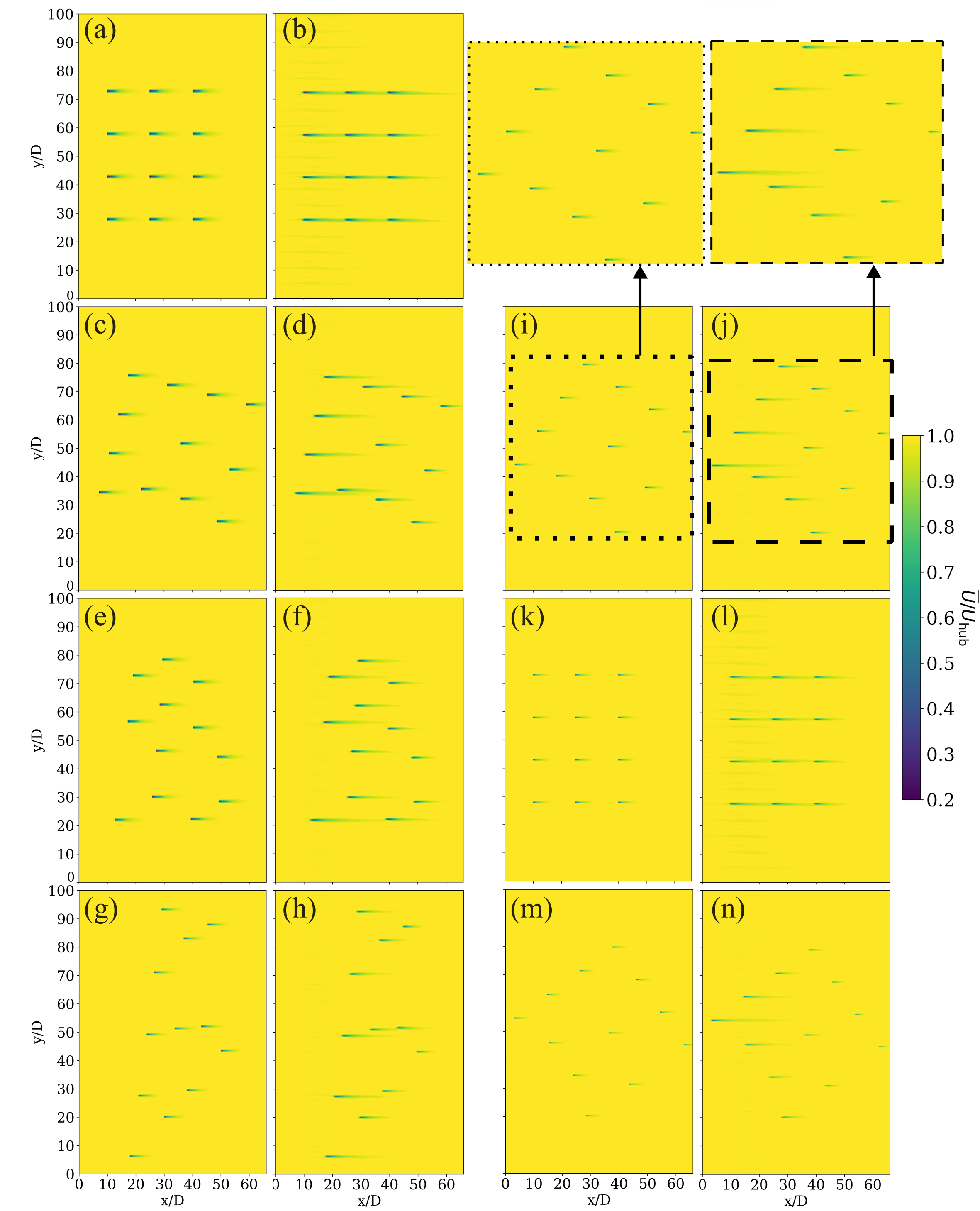}
  \caption{Color maps of wake flow results for seven training cases with unyawed turbines (see \prettyref{tab:0_2}). Panels (a) to (n) present top-view contour plots of the nondimensional time-averaged velocity magnitude $(\overline{U}/U_{\infty})$ at hub height. Each pair of subplots compares GCH model predictions (left) with LES results (right). Panels (a) and (b), (c) and (d), (e) and (f), (g) and (h), (i) and (j), (k) and (l), and (m) and (n) correspond to an inflow wind speed of 9, 10, 11, 13, 14, 15, and 17 m/s, respectively. Insets in (i) and (j) highlight regions of intensified wake interactions for denser array configurations. The wind direction is from left to right.}
  \label{Fig:2}
  \label{fig:2}
\end{figure}

\noindent \prettyref{fig:2} shows the time-averaged velocity field of the seven test cases with unyawed turbines at hub height elevation, computed using the LES and GCH model. As expected, a clear qualitative discrepancy is observed between the two models. The GCH wake model produces broad lateral spread and highly symmetric turbine wakes, which are characterized by high velocity deficit in the near-wake region of the turbines. This observation is consistent with previous findings reported by \citet{[25]} and \citet{[45]}. In contrast, the LES reveals more realistic wake dynamics with wakes that are slightly asymmetrical, and narrower. Also, as seen, the LES-computed wake flows could be characterized by lower momentum deficits in the near-wake region.

Additionally, the GCH model exhibits a faster wake recovery compared to LES results. This is aligned with previous findings by \citet{[25]}. Specifically, the wake recovery process in the GCH model is nearly completed within approximately $5D$ downstream of each turbine, whereas LES results show that wake recovery may extend up to $20D$ downstream. This discrepancy becomes particularly significant in scenarios where downstream turbines are influenced by the wakes of upstream turbines, known as wake-wake interaction~\cite{[169], [170]} (see, e.g., \prettyref{Fig:2}-a,b). Since the turbine power production scales with the cube of the incoming velocity ($\overline{U}^3$)~\cite{[25]}, the overestimation of wake recovery, as seen in the GCH model's results, could lead to higher flow velocities reaching downstream turbines and, thus, result in an over-prediction of available power.  Moreover, the wake flow results of both models show a decrease in wake deficit with increasing wind speed. As reported by \citet{[171]} and \citet{[172]}, this trend may be attributed to the reduction in thrust coefficient at higher wind speeds, which diminishes wake intensity. 

\begin{figure}[H]
  \includegraphics[width=0.95\textwidth]{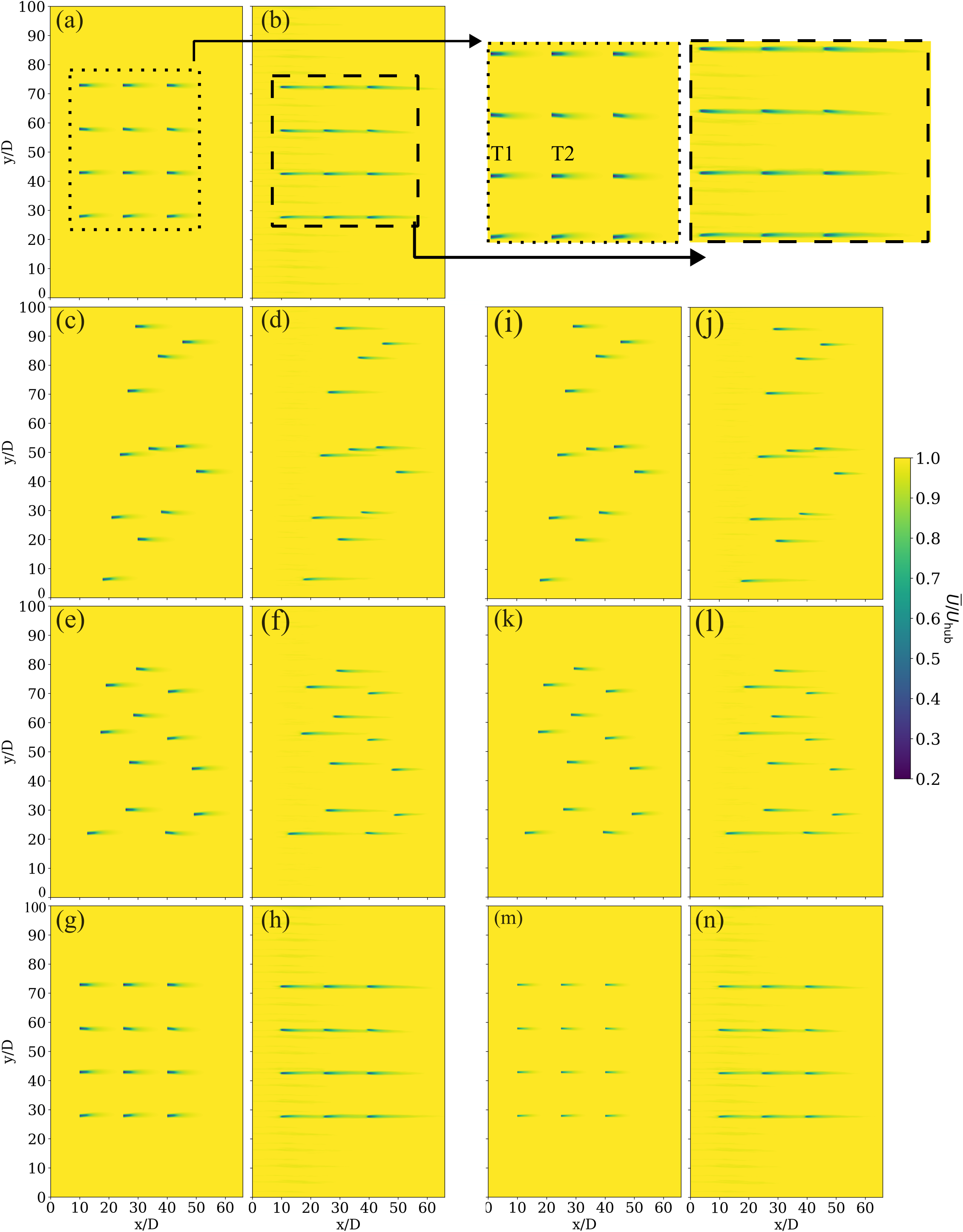}
  \caption{Color maps of wake flow results for seven training cases with yawed turbines (see \prettyref{tab:0_2}). Panels (a) to (n) present top-view contour plots of the nondimensional time-averaged velocity magnitude $(\overline{U}/U_{\infty})$ at hub height. Each pair of subplots compares the GCH model predictions (left) with the LES results (right). Panels (a) and (b), (c) and (d), (e) and (f), (g) and (h), (i) and (j), (k) and (l), and (m) and (n) correspond to an inflow wind speed of 5, 7, 7, 8, 10, 11, and 13 m/s, respectively. Insets in (a) and (b) highlight regions of intensified wake interactions for denser array configurations. The wind direction is from left to right.}
  \label{Fig:3}
\end{figure}

Concerning test cases with yawed turbines, as seen in \prettyref{Fig:3}, the GCH model captures the wake steering behavior of yawed turbines, though it slightly overpredicts the degree of wake deflection. In other words, the GCH model-computed wake is correctly skewed in the direction normal to the rotor plane, resulting from a pair of counter-rotating vortices that displace the wake, as previously demonstrated by \citet{[50]} and \citet{[25]}. 
Notwithstanding that the GCH model neglects the nacelle geometry and its impact on wake dynamics, it appears to successfully reproduce the main features of wake deflection of yawed turbines. This is consistent with the findings of \citet{[25]}. 

An interesting observation arises from the wake flow around the T1 and T2 turbines and their interactions in \prettyref{Fig:3}(a),(b). As described in \prettyref{Fig:3_1}, configuration 1, the T1 turbine is yawed while the T2 turbine is not.  For that, one would expect to see only one skewed wake past the T1 turbine. However, as seen in the LES results of \prettyref{Fig:3}(b), the wake steering is evident for the unyawed turbine, as well. In other words, the wake of the unyawed T2 turbine is also slightly deflected. This is known to be due to the influence of the upstream yawed turbine wake, i.e., the wake of T1 turbine. This wake behavior of waked turbines is referred to as the secondary wake steering~\cite{[50]}. We should note that while the GCH model is theoretically capable of capturing this effect~\cite{[25]}, its overly rapid wake recovery causes the upstream wake to dissipate before reaching the downstream turbine (\prettyref{Fig:3}(a)). Therefore, one could argue that the GCH model suppresses the secondary wake steering. Although such differences may seem minor, they become notably cumulative in wind farm configurations where downstream turbines are subjected to upstream turbine wakes. In such cases, the inability to capture the aforementioned wake dynamics may lead to inaccuracies in estimating the wind farm's power production. 

\begin{figure}[H]
  \includegraphics[width=1\textwidth]{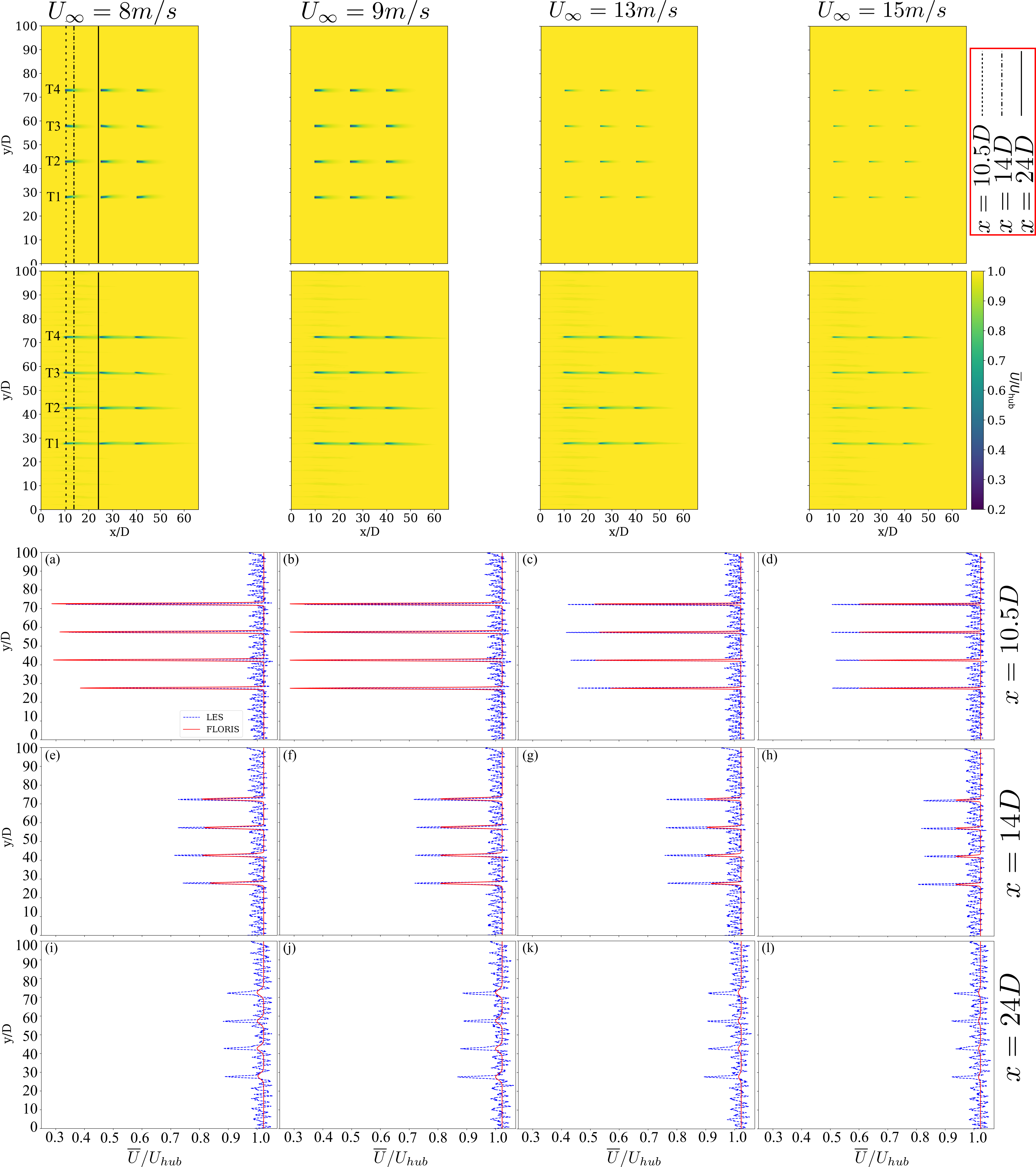}
  \caption{Comparison of cross-stream velocity profiles obtained from GCH model and LES at hub height. The first through fourth columns reffer to test cases $2$, $3$, $4$, and $5$, respectively. Panels (a–d), (e–h), and (i–l) show normalized mean velocity profiles $\overline{U}/U_{\mathrm{hub}}$ at $10.5D$, $14D$, and $24D$ downstream, respectively. Dashed, dash-dot, and solid vertical lines indicate those three downstream distances. Each subplot corresponds to a different inflow condition: (a,e,i) 8 m/s, (b,f,j) 9 m/s, (c,g,k) 13 m/s, and (d,h,l) 15 m/s. Blue dashed lines ({\color{blue} \dashed}) represent LES results, while solid red lines ({\color{red} \solid}) show GCH model predictions.}
  \label{Fig:4}
\end{figure}

To further examine the performance of the analytical model, in \prettyref{Fig:4} we compare profiles of velocity across test cases $2$, $3$, $4$, and $5$, encompassing a range of wind speeds under both yawed and unyawed conditions. The velocity profiles from LES (blue dashed line) and GCH model (red solid line) are obtained at hub height. Specifically, these profiles correspond to $x=10.5D$, $x=14D$, and $x=24D$ downstream of turbines, which represent positions immediately behind, moderately downstream, and far field, respectively. 
As seen in \prettyref{Fig:4}\textnormal{(a)} and \textnormal{(b)}, which correspond to the region immediately downstream of the turbines operating at wind speeds below the rated velocity, the GCH model overpredicts the wake deficit. Specifically, this overprediction, compared to LES, is calculated to be approximately \(6.7\%\) for the unyawed case and \(10.3\%\) for the yawed condition. This trend is reversed for wind speeds exceeding the rated velocity, as observed in \prettyref{Fig:4}(c), (d). In the near-wake region, the underprediction of the velocity deficit by GCH model compared to LES reaches approximately \(24.3\%\) for the unyawed case and \(20.6\%\) for the yawed condition. Similar findings were also reported by \citet{[25]} and \citet{[45]}.
Further downstream at $x=14D$, the GCH model results exhibit a much faster wake recovery compared to LES (\prettyref{Fig:4}(e)-(h)). At this location, the discrepancy between the GCH model and LES reaches an average of $14.1\%$ and $13.5\%$ for the unyawed and yawed turbines, respectively. Farther downstream at $x=24D$, the velocity predicted of the GCH model has nearly returned to freestream values, whereas the LES predictions still indicate residual wake effects, resulting in an average discrepancy of $8.7\%$ between GCH model and LES (\prettyref{Fig:4}(i)-(l)).
Moreover, as seen in the first and third columns of \prettyref{Fig:4} (i.e., cases $2$ and $4$), a slight underprediction of the velocity deficit is observed for the turbine operating under higher yaw angle condition (e.g., turbine T1) compared to those with lower yaw angles (e.g., turbines T2 and T4). This effect is captured by both the LES and the GCH model in the near-wake region (\prettyref{Fig:4}(a), (c)). However, at $x=14D$ downstream, the difference in velocity deficit between yawed and unyawed turbines becomes negligible in both the LES and the GCH model predictions.

To further investigate the differences between the GCH model and LES results, \prettyref{tab:1} presents four error metrics, including coefficient of determination ($R^2$), Mean Absolute Error (MAE), Root Mean Square Error (RMSE), and Mean Absolute Relative Error (MARE), to quantitatively evaluate the GCH-predicted flow field across the entire 3D domain. This comprehensive evaluation provides a more precise understanding of the GCH model performance. The definition of these error metrics are reported in ~\cite{[150]}. As shown in \prettyref{tab:1},
the GCH model exhibits a considerable level of error when compared to the LES results. Given that the GCH model can compute the velocity field within a few minutes for each case, it appears to be a practical and efficient tool for wind farm layout optimization. However, given the magnitude of the observed errors, the direct use of the GCH model may lead to significant inaccuracies in predicting the velocity field, and thus in estimating the wind farm's power production.

\begin{table}[H]
\centering
\caption{Error metrics for each test case based on the comparison between the GCH model and LES.}
\begin{tabular}{ccccc}
\textit{Test case} & \textit{\(R^2\)} & \textit{MAE} & \textit{RMSE} & \textit{MARE} \\
\midrule
1   & 0.7711 & 0.0908 & 0.1130 & 0.0850 \\
2   & 0.7713 & 0.0906 & 0.1129 & 0.0851 \\
3   & 0.7705 & 0.0909 & 0.1132 & 0.0854 \\
4   & 0.7768 & 0.0895 & 0.1115 & 0.0840 \\
5   & 0.7750 & 0.0900 & 0.1119 & 0.0840 \\
6   & 0.7765 & 0.0892 & 0.1113 & 0.0838 \\
7   & 0.7780 & 0.0891 & 0.1109 & 0.0840 \\
8   & 0.7842 & 0.0876 & 0.1091 & 0.0824 \\
9   & 0.7779 & 0.0888 & 0.1109 & 0.0833 \\
10  & 0.7808 & 0.0883 & 0.1102 & 0.0830 \\
11  & 0.7809 & 0.0883 & 0.1102 & 0.0829 \\
12  & 0.7795 & 0.0884 & 0.1105 & 0.0830 \\
13  & 0.7857 & 0.0870 & 0.1087 & 0.0820 \\
14  & 0.7866 & 0.0868 & 0.1084 & 0.0812 \\
\end{tabular}
\label{tab:1}
\end{table}

\subsection{Robustness of the ML model}
\noindent The ML model was meticulously trained on the datasets obtained from LES in test cases $1$ through $14$. While the robustness of the model had been previously established for smaller turbine arrays~\cite{[25]}, herein, we extend its application to a utility-scale wind farm consisting of $12$ turbines. The generalization capability of the trained ML model will be evaluated against the LES results of the test case $15$, i.e., a previously unseen configuration.

\begin{figure} [H]
  \includegraphics[width=1\textwidth]{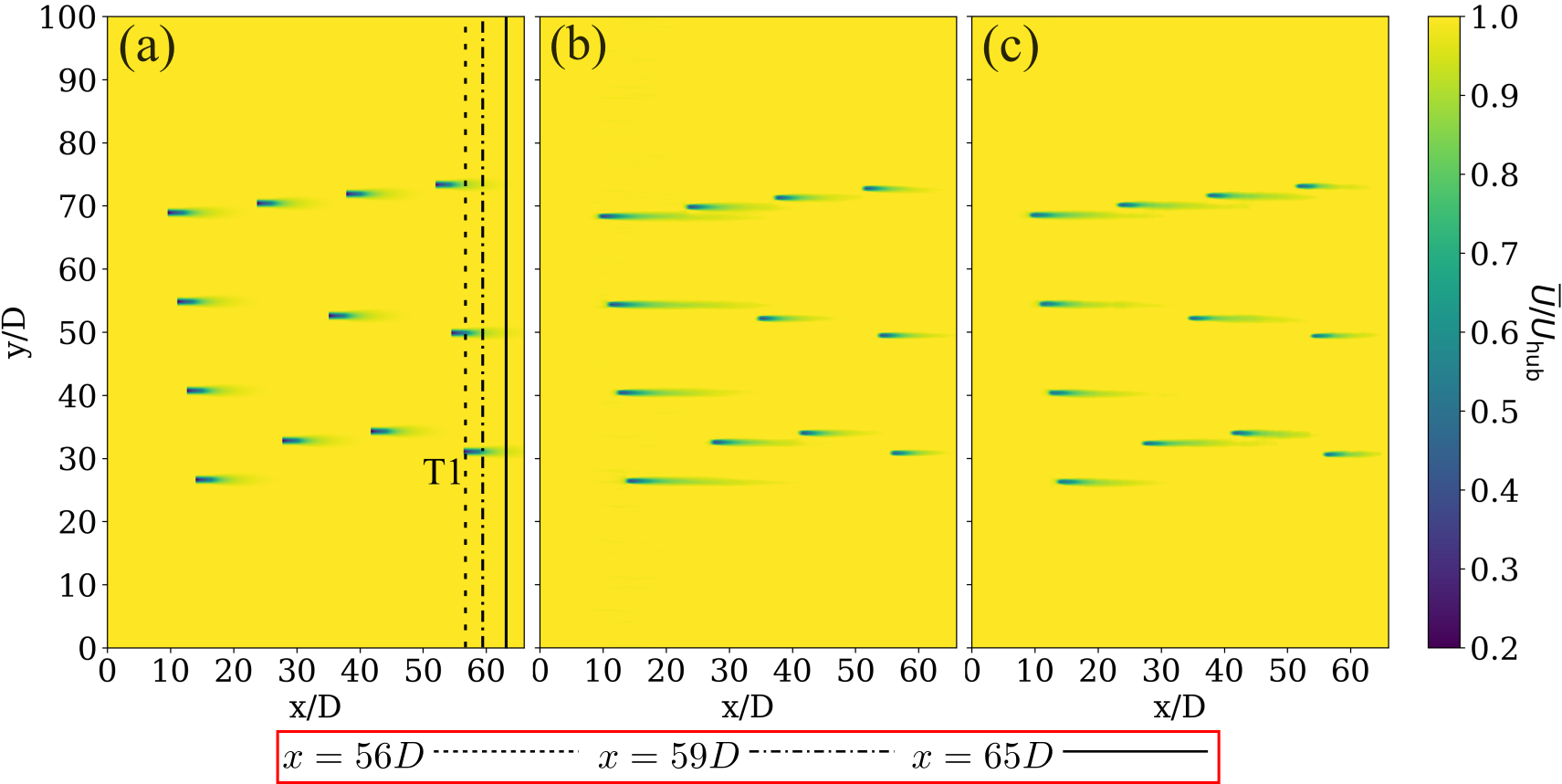}
  \caption{Color maps of wake flow results for one testing case with unyawed turbines (see \prettyref{tab:0_2}). Panels (a) to (c) present top-view contour plots of the nondimensional time-averaged velocity magnitude $(\overline{U}/U_{\infty})$ at hub height. Each pair of subplots compares the GCH model predictions (left), the LES results (middle), and the ML model predictions (right). The dashed, dash-dot, and solid vertical lines in (a) denote distances of $x=56D, 59D,$ and $65D$ downstream, respectively.
  The inflow wind speed is $7 \ m/s$, with the wind direction from left to right.}
  \label{Fig:5_1}
\end{figure}

\begin{figure} [H]
  \includegraphics[width=1\textwidth]{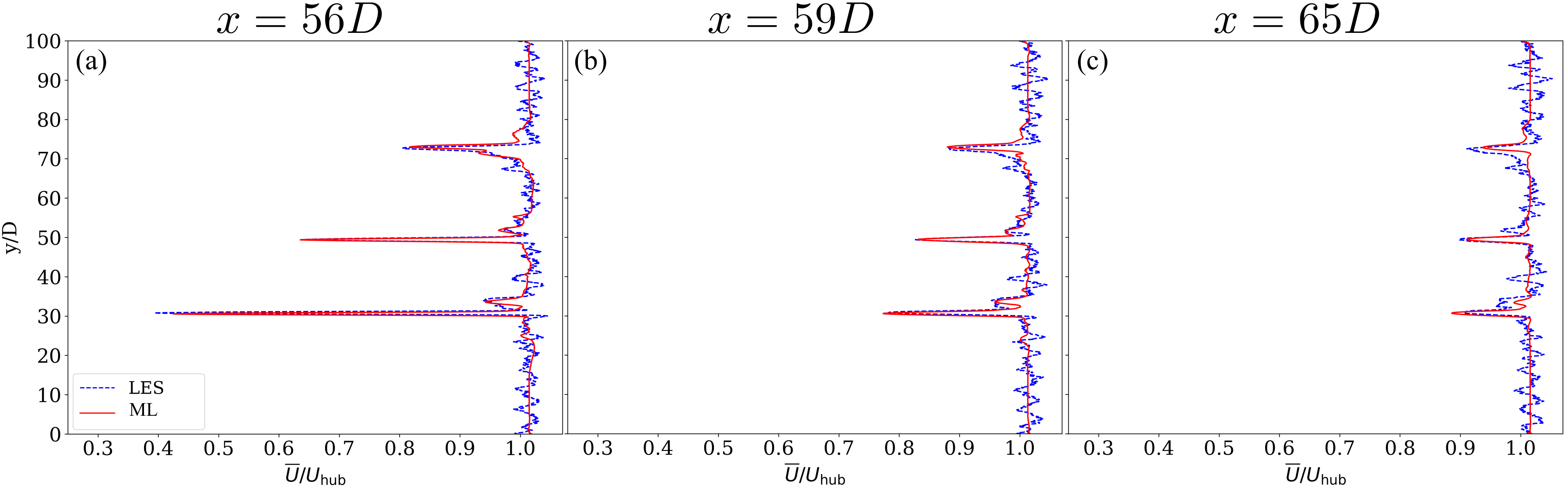}
  \caption{Cross-stream velocity profiles at hub height predicted by LES and the ML model. Panels (a) to (c) show normalized mean velocity profiles $\overline{U}/U_{\mathrm{hub}}$ at $56D$, $60D$, and $65D$ downstream, respectively (see \prettyref{Fig:5_1} (a)). Blue dashed lines ({\color{blue} \dashed}) represent LES results, while solid red lines ({\color{red} \solid}) indicate predictions from the LES–ML model. The figure illustrates the model ability to accurately capture wake structures and recovery trends at far-wake locations.}
  \label{Fig:5_2}
\end{figure}

In \prettyref{Fig:5_1}, we present the time-averaged velocity field at the hub height of the turbines obtained from the GCH model, LES, and the ML model. Overall, the ML model shows excellent agreement with the LES predictions. Both the near- and far-wake structures are well captured by the ML model. Moreover, the wake flow velocity profiles of different models are plotted in \prettyref{Fig:5_2}. As seen in \prettyref{Fig:5_2}(a) to (c), the ML predictions exhibit excellent agreement with the LES results in both the near and far wake regions. Importantly, comparing the velocity profiles from the LES and the ML model, the velocity deficit and the wake recovery process are predicted with relatively negligible error.

In \prettyref{tab:2}, we present the four error metrics, including \(R^2\), MAE, RMSE, and MARE, of the ML and GCH models,  assesing the ML-predicted flow field across the entire wind farm. As seen, the ML model exhibits great agreement with the LES results across all error measures, especially in comparison to the test case predicted by the GCH model: the coefficient of determination $R^2$ increases from $0.778$ to $0.99$; the MAE decreases from $0.0887$ to $0.0159$; the RMSE decreases from $0.1109$ to $0.0234$; and the MARE decreases from $0.0833$ to $0.0177$. Due to the substantial reduction in error metrics and the potential for improved accuracy in AEP prediction, the ML model demonstrates strong reliability and is therefore well-suited for use in subsequent wind farm layout optimization.

\begin{table}[H]
\centering
\caption{Error metrics for the unseen test case $15$ to compare the ML and GCH models' wake flow predictions against LES results.}
\vspace{0.3em}
\renewcommand{\arraystretch}{1.2}
\begin{tabular}{@{}lccccc@{}}
\textit{Test case} & \textit{Prediction Model} & \textit{${R^2}$} & \textit{MAE} & \textit{RMSE} & \textit{MARE} \\
\midrule
\multirow{2}{*}{15} & ML     & 0.9901 & 0.0159 & 0.0234 & 0.0177 \\
                     & GCH & 0.7781 & 0.0887 & 0.1109 & 0.0833 \\
\end{tabular}
\label{tab:2}
\end{table}

\subsection{Wind farm optimization procedure}
\begin{figure} [H]
 \includegraphics[width=1.0\textwidth]{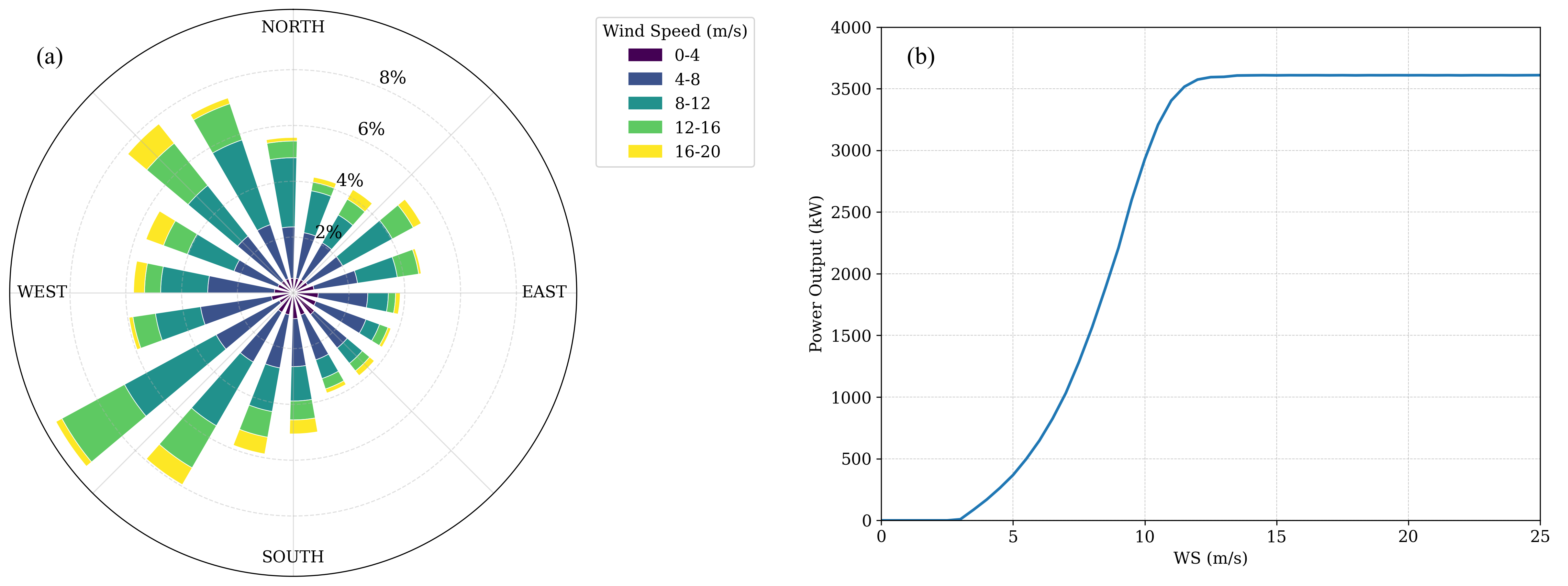}
  \caption{On-site wind rose diagram representing the wind speed and direction distribution at the South Fork wind farm for the year 2020 \cite{[174]} (a). The power curve of the Siemens Gamesa SWT-3.6-120 wind turbine as a function of effective wind speed (b).}
  \label{Fig:8}
\end{figure}

\noindent The optimization aims to maximize AEP by strategically positioning turbines within a defined spatial domain. This is done by accounting for varying wind speeds and directions (local wind rose) to minimize wake loss and increase available power at downstream turbines, and ultimately improve the wind farm's overall power production. This is particularly important in large-scale wind farms, where suboptimal layout decisions can significantly impact performance over the farm's operational lifespan. The annual energy production is calculated as~\cite{[173]}:
\begin{equation}
AEP = H \cdot \sum_{i=1}^{M} \sum_{j=1}^{N} \left( \sum_{k=1}^{T} P_{i,j,k} \right) \cdot p_j \cdot p_i,
\label{eq:24}
\end{equation}
where $p_i$ and $p_j$ are the probability density functions of the wind direction ($i$) and wind speed ($j$), respectively. $H$ is the total number of turbine operating hours per year (=$8760h$). $M$ and $N$ are the total number of wind direction and wind speed bins, respectively. $T$ denotes the total number of turbines within the wind farm ($=12$). The probability density functions are derived from on-site measurements (through Wind Integration National Dataset Toolkit~\cite{[174]}) and visualized using a wind rose diagram to illustrate the predominant wind directions and speeds (\prettyref{Fig:8}(a)). $P$ denotes the single turbine power output, derived from the power curve as a function of wind speed (WS), as shown in \prettyref{Fig:8}(b). The wind speed is defined as the mean velocity, averaged over the rotor swept area, and calculated as follows:
\begin{equation}
WS = \frac{\int_{A_{rot}} \overline{U} \, dA}{A_{rot}}.
\label{eq:25}
\end{equation}
where $A_{rot}$ is the rotor swept area ($=\pi D^2/4$), and $\overline{U}$ is the time-averaged velocity obtained from the ML model.

The implemented optimization approach is based on the greedy algorithm. At each step, greedy algorithm 
makes locally optimal choice with the expectation that it will lead to a globally optimal solution. In other words, during the optimization process, the algorithm iteratively adjusts the decision variables so that they yield the most immediate improvement in the objective function \cite{[155], [156]}. The configuration yielding the highest AEP in each iteration is retained for subsequent steps. Although this algorithm does not guarantee convergence to a global optimum, it is frequently employed in large-scale optimization problems due to its simplicity, computational efficiency, and ability to yield reasonably good solutions in practice~\cite{[65], [66]}. The procedure of this algorithm is detailed in Algorithm~\ref{alg:greedy}. 

\begin{algorithm}
\vspace{10pt}
\caption{Greedy Algorithm for Annual Energy Production Optimization}
\label{alg:greedy}
\begin{algorithmic}[1]
\Require Number of turbines $N_t$, positions per turbine $N_x$, position increment $\Delta x_j$
\Ensure Optimal coordinates $x_i$, maximum AEP

\State \textbf{Initialize:} $x_i \gets$ Random positions for all turbines
\State \textbf{Initialize:} Preliminary AEP $\gets$ \textbf{PAEP}
\For{$i = 1:N_t$} \Comment{Loop through turbines}
    \For{$j = 1:N_x$} \Comment{Explore position options}
        \State $x_i \gets x_i + j \Delta x_j$
        \State \textbf{AEP} $\gets f(x_i)$
        \If{\textbf{AEP} $>$ \textbf{PAEP}}
            \State \textbf{PAEP} $\gets$ \textbf{AEP}
            \State $y_i \gets x_i$
        \EndIf
    \EndFor
    \State $x_i \gets y_i$
\EndFor
\Statex \textbf{Output:} Maximum $AEP = f(x_i)$
\end{algorithmic}
\end{algorithm}

\begin{figure}[H]
\caption{Flowchart of the optimization framework illustrating the coupling of the greedy algorithm with FLORIS and the ML model.}
\begin{tikzpicture}[node distance=1.5cm and 2.5cm]

\tikzstyle{startstop} = [rectangle, rounded corners, minimum width=3.5cm, minimum height=1cm, text centered, draw=black]
\tikzstyle{process} = [rectangle, minimum width=3.5cm, minimum height=1cm, text centered, draw=black]
\tikzstyle{decision} = [diamond, aspect=2, text centered, draw=black]
\tikzstyle{io} = [trapezium, trapezium left angle=70, trapezium right angle=110, minimum width=3.5cm, minimum height=1cm, text width=3.5cm, align=center, draw=black]
\tikzstyle{arrow} = [thick,->,>=stealth]

\node (start) [startstop] {Read configuration file};
\node (hist) [process, below of=start] {Read histogram of site wind speed and direction};
\node (genvel) [process, below of=hist] {FLORIS-GCH: Generate low-fidelity velocity fields};
\node (enhance) [process, below of=genvel] {ML-model: Generate high-fidelity velocity fields};
\node (aep) [process, below of=enhance] {Compute AEP};
\node (compare) [decision, below of=aep, yshift=-0.5cm] {AEP $>$ AEP$_{\text{old}}$?};

\node (update) [process, right=3cm of genvel] {Update turbine placement};

\node (savecoords) [process, below of=compare, yshift=-0.5cm] {Save turbine coordinates};
\node (jcheck) [decision, below of=savecoords, yshift=-0.5cm] {$j > N_x$?};
\node (icheck) [decision, below of=jcheck, yshift=-0.5cm] {$i > N_t$?};
\node (finalsave) [io, below of=icheck, yshift=-0.5cm] {Save the optimal turbine layout and AEP};

\draw [arrow] (start) -- (hist);
\draw [arrow] (hist) -- (genvel);
\draw [arrow] (genvel) -- (enhance);
\draw [arrow] (enhance) -- (aep);
\draw [arrow] (aep) -- (compare);
\draw [arrow] (compare) -- node[anchor=north east, pos=0.6, yshift=10pt] {Yes} (savecoords);
\draw [arrow] (savecoords) -- (jcheck);

\draw [arrow] (jcheck) -- node[anchor=north east, pos=0.6, yshift=10pt] {Yes} (icheck);
\draw [arrow] (jcheck.east) -| node[anchor=south west, pos=0.25] {No, $j++$} (update.south);

\draw [arrow] (icheck) -- node[anchor=north east, pos=0.6, yshift=12pt] {Yes} (finalsave);
\draw [arrow] (icheck.east) -| node[anchor=south west, pos=0.25] {No, $i++$} (update.south);

\draw [arrow] (update.west) -- ++(-2.25,0) |- (genvel.east);
\draw [arrow] (compare.west) -- ++(-2,0) node[anchor=north east, near start, yshift=0pt] {No} |- (jcheck.west);

\end{tikzpicture}
\label{Fig:5}
\label{fig:5}
\end{figure}
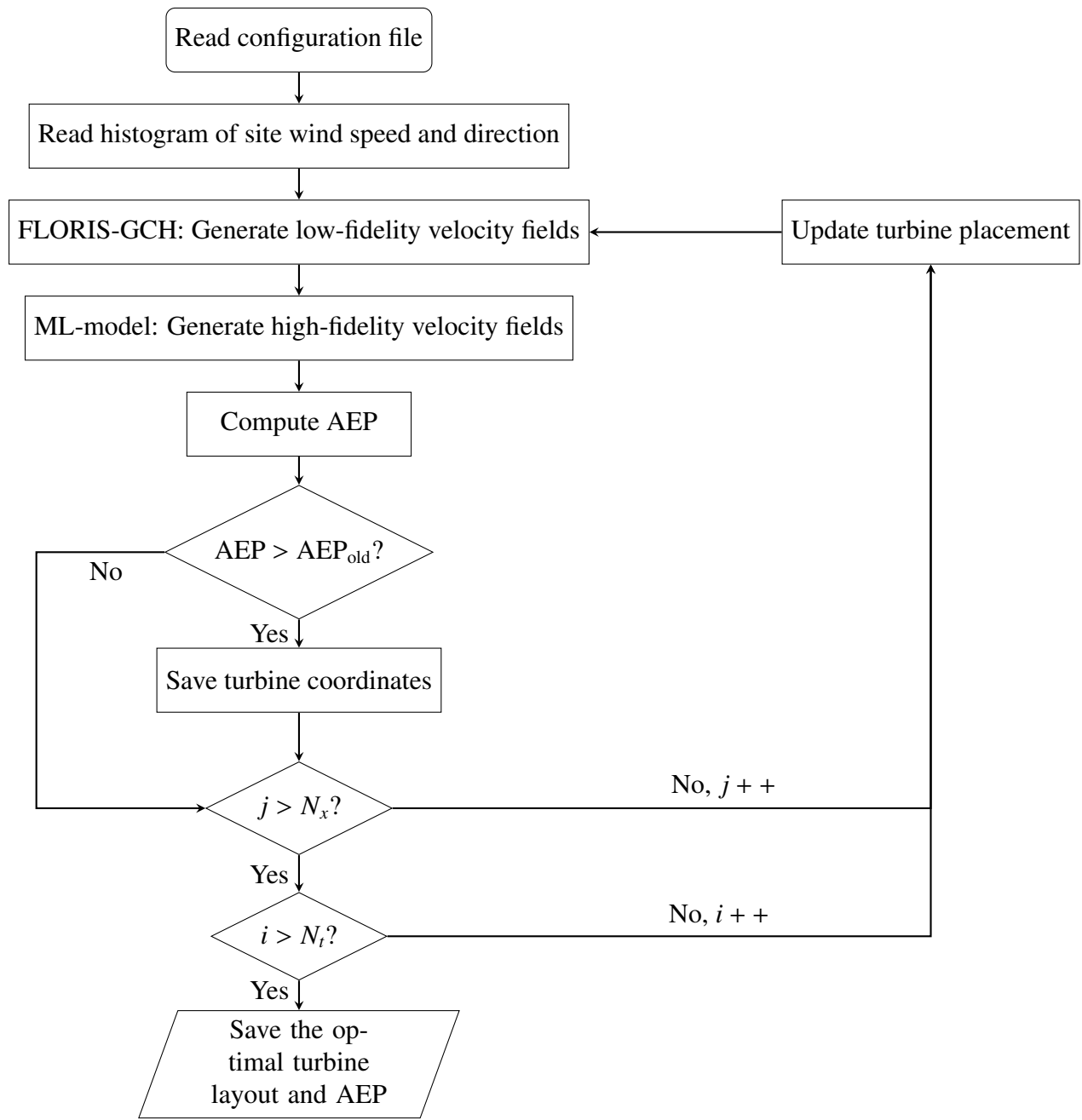

After having described the greedy algorithm function, now we elucidate its implementation in our optimization process. The procedure begins by reading the input data; including wind farm settings, turbine configuration, and site-specific wind speed and direction data (see \prettyref{Fig:8}(a)); to compute the AEP. The computational domain is first discretized into a $5 \times 5$ grid for turbine positioning. Upon randomly distributing the turbines within these grid cells, the GCH model obtains the low-fidelity velocity field for the given layout. This velocity field then serves as the input to the trained ML model, which generates high-fidelity velocity field predictions. Optimization takes place according to the greedy algorithm: turbine locations are updated one at a time within the unoccupied grid cells; and for each modified layout, the AEP is recalculated (\prettyref{eq:24}) using the high-fidelity velocity field predicted by the ML model. The configuration that yields the higher AEP compared to the previous iteration is retained. Once all turbines and feasible non-occupied positions have been evaluated, the layout that results in the maximum AEP is selected as the optimal configuration. \prettyref{fig:5} illustrates the entire optimization process, which entails the greedy algorithm with the GCH model and the ML model. Notably, the flow domain is rotated for each case such that the streamwise axis aligns with the corresponding wind direction, assuming an unyawed turbine configuration. For further details regarding the domain rotation methodology, the reader is referred to \cite{[62]}.

\subsection{Optimal wind farm layout}
\noindent We employed the above-described optimization algorithm using the ML model to obtain a wind farm layout with maximal AEP without increasing land use. Also, the ML model was used to obtain the wake flow of the baseline case, i.e., existing layout of the South Fork wind farm, under various wind directions and wind speeds. The results were subsequently used to establish a baseline for the calculation of AEP, and to compare the performance of the baseline wind farm with the optimized wind farm obtained from the greedy optimization algorithm. The comparisons are carried out under wind directions of $10^\circ$, $230^\circ$, and $330^\circ$ corresponding to wind speeds of $5$ m/s, $9 $ m/s, and $15$ m/s, respectively. This selection is to evaluate the performance of the optimized layout under varying conditions, including one sub-dominant wind (wind direction of $10^\circ$, wind speed of $5$ m/s) and two dominant scenarios (wind direction of $230^\circ$ at $9 \ m/s$ and $330^\circ$ at $15 \ m/s$).

\begin{figure} [H]
 \includegraphics[width=1\textwidth]{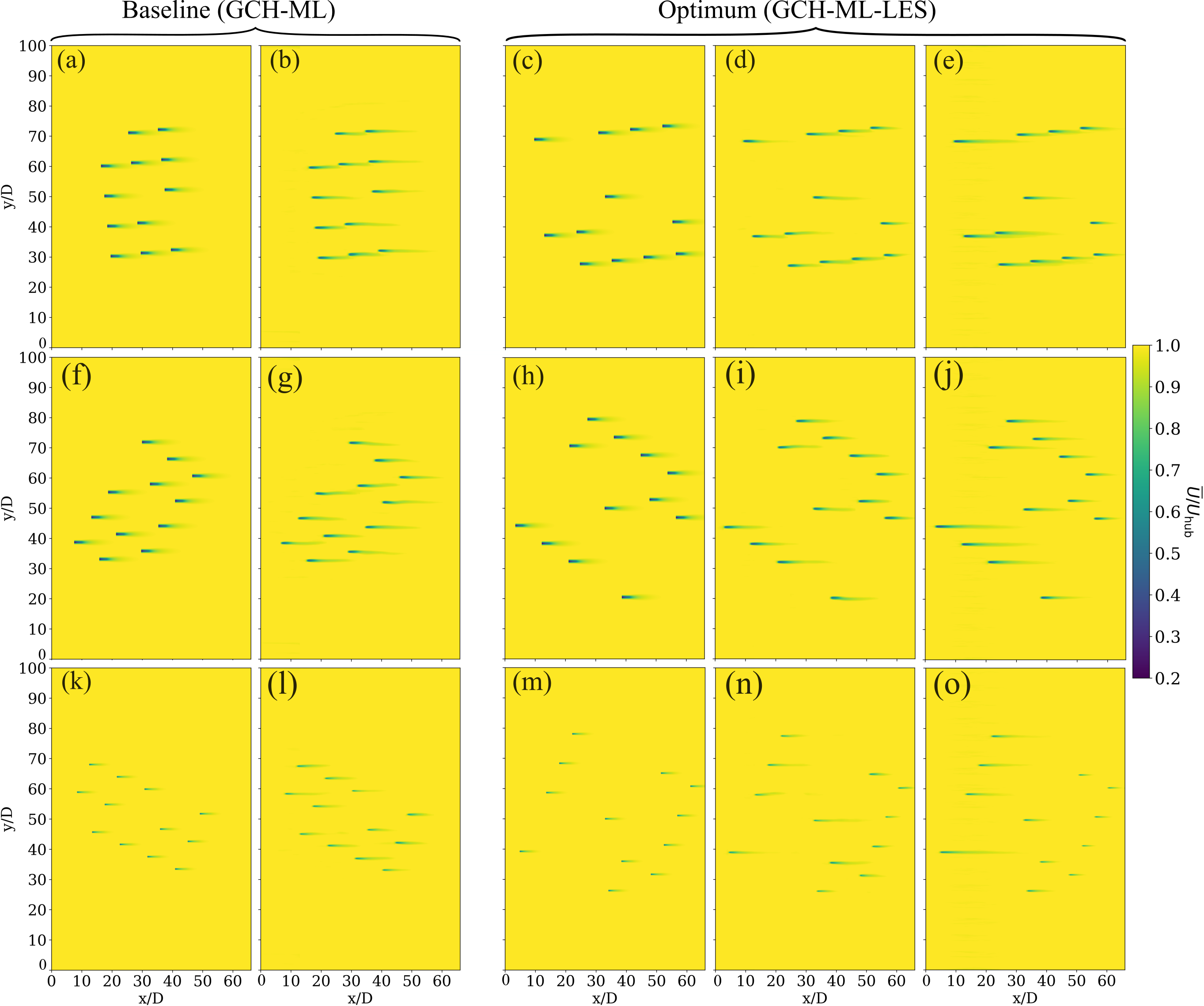}
  \caption{Color maps of wake flow results for the baseline and optimized configurations under varying wind conditions. Panels (a) to (o) present top-view contour plots of the nondimensional time-averaged velocity magnitude $(\overline{U}/U_{\infty})$ at hub height.
  Panels (a) and (b), (f) and (g), and (k) and (l) represent the baseline configurations predicted by the GCH (left), and the ML (right) model, while panels (c–e), (h–j), and (m–o) show the corresponding optimized configurations determined by the GCH (left), the ML model (middle), and the LES (right). Panels (a-e), corresponds to a wind direction of 10$^{\circ}$ and wind speed of 5 m/s, panels (f-j) corresponds to a wind direction of 230$^{\circ}$ and wind speed of 9 m/s, and panels (k-o) corresponds to a wind direction of 310$^{\circ}$ and wind speed of 15 m/s. The wind direction is from left to right.}
  \label{Fig:6}
\end{figure}

We plot in \prettyref{Fig:6} the low- and high-fidelity time-averaged velocity fields at hub height, predicted by GCH model and ML for the baseline layout, and by GCH model, ML, and LES for the optimized layout, under both sub-dominant and dominant scenarios. As seen, in the optimized configuration, the turbine distribution is slightly more spread in the spanwise direction and considerably more stretched along the streamwise axis. This arrangement for the optimal configuration aligns with previous findings by \citet{[138]}, which suggest that for a given turbine density, increasing streamwise spacing leads to greater performance improvements than increasing spanwise separation. While some turbines in the optimized layout remain aligned similarly to the baseline (\prettyref{Fig:6}(c)-(e)), the adjustments introduced by the optimization algorithm remain beneficial, as this particular wind condition is classified as a sub-dominant scenario and thus contributes less significantly to the overall AEP.

In \prettyref{Fig:6}(f)–(o), under dominant wind conditions (i.e., wind directions of $230^\circ$ and $330^\circ$), the optimized layout features expanded turbine spacing in both the spanwise and streamwise directions compared to the baseline configuration. This arrangement ensures that nearly all turbines avoid significant wake interference from adjacent and upstream units, ultimately resulting in increased overall energy output due to reduced wake losses.

Quantitatively, the AEP across all considered conditions is calculated as $140.52 GWh$ and $143.53 GWh$ for the baseline and optimized layout, respectively, marking a $2.05 \%$ increase in AEP of the optimal layout. Although the improvement may appear modest, it becomes highly significant over a typical wind farm operational lifespan of approximately $25$ years \cite{[160]}, resulting in an estimated $50\%$ cumulative gain in total energy yield over the project lifetime.


We now concentrate on evaluating the performance of the ML model in predicting the high-fidelity velocity field for the optimized turbine layout. For brevity, we compare the ML-predicted flow fields against the LES results only for the representative wind directions and speeds discussed above. As seen in \prettyref{Fig:6}(e), (j), and (o), the ML model demonstrates strong agreement with the LES across all scenarios, accurately capturing key features such as wake–wake interactions, velocity deficit, and wake recovery process. Furthermore, \prettyref{tab:3} provides a quantitative error analysis by comparing the ML and GCH model predictions against LES for the optimized layout. The results clearly indicate that the ML model closely matches the LES data, significantly outperforming the GCH model. Along with the approximately $950\,000$-fold reduction in computational time compared to the LES, these findings confirm that the optimization process is robust, reliable, and cost-effective. As a result, the optimized turbine layout and its associated AEP can be considered valid for practical implementation.

\begin{table}[H]
\centering
\caption{Error metrics for the optimum layout, comparing ML and GCH models against LES across different wind conditions.}
\vspace{0.3em}
\renewcommand{\arraystretch}{1.2}
\begin{tabular}{@{}l c c c c c c@{}}
\textit{Wind Speed} & \textit{Wind Direction} & \textit{Prediction Model} & \textit{$R^2$} & \textit{MAE} & \textit{RMSE} & \textit{MARE} \\
\midrule
\multirow{2}{*}{5 $m/s$} & \multirow{2}{*}{$10^\circ$}   & ML      & 0.9867 & 0.0200 & 0.0273 & 0.0219 \\
                       &                               & GCH  & 0.7725 & 0.0906 & 0.1128 & 0.0857 \\
\addlinespace
\multirow{2}{*}{9 $m/s$} & \multirow{2}{*}{$230^\circ$}  & ML      & 0.9839 & 0.0218 & 0.0300 & 0.0235 \\
                       &                               & GCH  & 0.7693 & 0.0911 & 0.1137 & 0.0854 \\
\addlinespace
\multirow{2}{*}{15 $m/s$} & \multirow{2}{*}{$310^\circ$} & ML      & 0.9841 & 0.0221 & 0.0297 & 0.0239 \\
                       &                               & GCH & 0.7784 & 0.0894 & 0.1110 & 0.0838 \\
\end{tabular}
\label{tab:3}
\end{table}

\section{Conclusion}
\label{sec:5}

\noindent Optimizing wind farm layouts remains challenging due to the high computational cost of high-fidelity simulations. While reduced-order models such as GCH are more efficient, their deviation from LES results limits their effectiveness for layout optimization. This study seeks to integrate an ML model, trained with high-fidelity LES data, with GCH model for efficient wake flow predictions and, thus, optimization of wind farm layouts. 

First, we evaluated the performance of the GCH model for velocity field predictions in a wind farm of $12$ turbines with varying turbine layouts, wind speeds, and yaw conditions. Our findings indicate that the FLORIS-implemented GCH model overestimates near wake velocity deficits at sub-rated wind speeds and underestimates them at super-rated wind speeds. Additionally, the wake recovery prediction of the GCH model is significantly faster than that observed in the LES results. While the GCH model is theoretically capable of capturing wake deflection and secondary wake steering effects~\cite{[25]}, its overly rapid wake recovery leads to dissipation of the upstream wake before it reaches the downstream turbine, thereby suppressing secondary steering. These limitations suggest that the GCH model alone may not offer the high level of accuracy required for reliable wind farm layout optimization.

To improve the wake flow predictions and enable a more accurate optimization process, we developed a machine learning model to map the low-fidelity velocity fields generated by the GCH model to high-fidelity counterparts. The proposed ML framework employs an ACNN with U-Net-style skip connections, which is trained and fine-tuned using LES-generated high-fidelity data. The model was trained by low-fidelity velocity fields from GCH model as input and by LES results as target. When tested on an unseen turbine layout, the trained ML model significantly improved prediction accuracy in key metrics such as velocity deficit and wake recovery, reducing the GCH model prediction error from over $11 \%$ to below $3\%$ across the wind farm domain.

The trained ML model was subsequently integrated into a wind farm layout optimization framework, which is based on a greedy algorithm. The South Fork wind farm served as the case study with $12$ turbines. The optimization process begins with the GCH model generating low-fidelity velocity fields for randomly distributed turbines within the computational domain, which are subsequently refined by the ML model to predict the high-fidelity flow field. The wake flow predictions were then used to compute the AEP. Afterwards, the turbine positions are iteratively updated within the unoccupied grid cells to locally maximize AEP. Although the greedy algorithm does not guarantee a globally optimal solution, it is a practical choice for large-scale wind farm optimization due to its simplicity, computational efficiency, and demonstrated success in prior studies~\cite{[65], [66]}.

The accuracy of the ML model in predicting the velocity field of the optimized layout was further assessed through comparison with the LES results. This evaluation demonstrated strong agreement with the LES and notable improvements over the GCH model, thereby confirming the reliability of the optimization process. When compared to the baseline configuration, the optimized layout achieved a $2.05\%$ increase in AEP, which translates to an estimated $50\%$ cumulative increase in energy yield over the typical operational lifetime of the wind farm~\cite{[160]}. Given the higher accuracy of the ML model compared to the GCH model predictions and its approximately $950\,000$-fold reduction in computational cost relative to the LES, the proposed ML-based optimization framework offers a reliable and cost-effective solution for wind farm layout optimization.

As a potential direction for future work, we propose extending the optimization framework to consider not only turbine performance but also structural aspects, aiming to minimize the levelized cost of energy~\cite{[151], [152], [153]}.

\section*{Acknowledgements}
\label{sec:acknowledge}
\noindent This work was supported by grants from the U.S. Department of Energy’s Office of Energy Efficiency and Renewable Energy (EERE) under the Water Power Technologies Office (WPTO) Award Numbers DE-EE0009450 and DE-EE00011379. Partial support was provided by NSF (grant number 2233986). The computational resources for the simulations of this study were partially provided by the Institute for Advanced Computational Science at Stony Brook University. The views expressed herein do not necessarily represent the views of the U.S. Department of Energy or the United States Government.

\section*{Author Contributions}
\noindent
\textbf{Mehrshad Gholami Anjiraki:} Conceptualization (equal); Data curation (equal); Formal analysis (equal); Investigation (equal); Methodology (equal); Visualization (equal); Writing – original draft (equal); Writing – review \& editing (equal). \textbf{Christian Santoni:} Conceptualization (equal); Data curation (equal); Investigation (equal); Methodology (equal); Writing – review \& editing (equal). \textbf{Samin Shapourmiandouab:} Investigation (equal); Visualization (equal); Writing – review \& editing (equal). \textbf{Hossein Seyedzadeh:} Investigation (equal); Visualization (equal); Methodology (equal); Writing – review \& editing (equal). \textbf{Jonathan Craig:} Conceptualization (equal); Writing – original draft (equal); Writing – review \& editing (equal). \textbf{Ali Khosronejad:} Conceptualization (equal); Data curation (equal); Formal analysis (equal); Funding acquisition (lead); Investigation (equal); Methodology (equal); Project administration (lead); Resources (lead); Software (lead); Supervision (lead); Visualization (equal); Writing – original draft (equal); Writing – review \& editing (equal).


\section*{Data Availability Statement}
\label{sec:7}
\noindent The software code (VFS-3.4 model) \href{ https://doi.org/10.5281/zenodo.16460624}{(10.5281/zenodo.16460624)}, along with the LES-generated flow field results \href{https://doi.org/10.5281/zenodo.16459499}{(10.5281/zenodo.16459499)}, the GCH model predictions \href{https://doi.org/10.5281/zenodo.16459134}{(10.5281/zenodo.16459134)}, the ML model predictions \href{https://doi.org/10.5281/zenodo.16459939}{(10.5281/zenodo.16459939)}, the ML-based optimization code, the trained ML model \href{https://doi.org/10.5281/zenodo.16460628}{(10.5281/zenodo.16460628)}, the AL segments, and nacelle surface files \href{https://doi.org/10.5281/zenodo.16460646}{(10.5281/zenodo.16460646)} are all made freely available via Zenodo online repository.

\clearpage
\appendix
\section{Blockwise Architecture of the U-Net Model}
\label{appendix:app1}
We present in \prettyref{tab:A.7} a detailed blockwise specification of the proposed U-Net model architecture, including input/output shapes, operation types, and layer configurations used during training and inference.

\begin{landscape}
\clearpage

\vspace{1em}

\renewcommand{\arraystretch}{1.1}  
\setlength{\tabcolsep}{2pt}        

\begin{table}[h!]
\centering
\begin{adjustbox}{width=1.02\textwidth,center}
\begin{tabular}{|
  >{\centering\arraybackslash}m{2.5cm}|
  >{\centering\arraybackslash}m{4cm}|
  >{\centering\arraybackslash}m{4.5cm}|
  >{\centering\arraybackslash}m{9cm}|
  >{\centering\arraybackslash}m{3cm}|}
\hline
\textit{Block} & \textit{Input Shape} & \textit{Operation Name} & \textit{Operation Details} & \textit{Output Shape} \\
\hline
InConv & $(B,\,79,\,L,\,W)$ & Double Conv2D & Conv2D $79\to64$, ReLU $\times2$ & $(B,\,64,\,L,\,W)$ \\
\hline
Down1 & $(B,\,64,\,L,\,W)$ & Downsample + Double Conv2D & MaxPool $\to \tfrac{L}{2}\times\tfrac{W}{2}$; Conv2D $64\to128$, ReLU $\times2$, Dropout(0.1) & $(B,\,128,\,\tfrac{L}{2},\,\tfrac{W}{2})$ \\
\hline
Down2 & $(B,\,128,\,\tfrac{L}{2},\,\tfrac{W}{2})$ & Downsample + Double Conv2D & MaxPool $\to \tfrac{L}{4}\times\tfrac{W}{4}$; Conv2D $128\to256$, ReLU $\times2$, Dropout(0.1) & $(B,\,256,\,\tfrac{L}{4},\,\tfrac{W}{4})$ \\
\hline
Down3 & $(B,\,256,\,\tfrac{L}{4},\,\tfrac{W}{4})$ & Downsample + Double Conv2D & MaxPool $\to \tfrac{L}{8}\times\tfrac{W}{8}$; Conv2D $256\to512$, ReLU $\times2$, Dropout(0.1) & $(B,\,512,\,\tfrac{L}{8},\,\tfrac{W}{8})$ \\
\hline
Down4 & $(B,\,512,\,\tfrac{L}{8},\,\tfrac{W}{8})$ & Downsample + Double Conv2D & MaxPool $\to \tfrac{L}{16}\times\tfrac{W}{16}$; Conv2D $512\to512$, ReLU $\times2$, Dropout(0.1) & $(B,\,512,\,\tfrac{L}{16},\,\tfrac{W}{16})$ \\
\hline
Parameter & $(B,\,5)$ & MLP + Transformer + Broadcast & MLP $\to$ Transformer $\to$ Broadcast & $(B,\,32,\,\tfrac{L}{16},\,\tfrac{W}{16})$ \\
\hline
Bottleneck &
\begin{tabular}{@{}c@{}}
$(B,\,512,\,\tfrac{L}{16},\,\tfrac{W}{16}) +$\\
$(B,\,32,\,\tfrac{L}{16},\,\tfrac{W}{16})$
\end{tabular} 
& Concat + 1$\times$1 Conv + Residual 
& Concat $\to$ 1$\times$1 Conv $544\to512$, ReLU $\to$ Residual add 
& $(B,\,512,\,\tfrac{L}{16},\,\tfrac{W}{16})$ \\
\hline
Up1 & $(B,\,512,\,\tfrac{L}{16},\,\tfrac{W}{16})$ 
& Upsample + Double Conv2D 
& ConvTranspose2D $512\to512$; Concat with $(B,\,512,\,\tfrac{L}{8},\,\tfrac{W}{8})$; Conv2D $1024\to512$, ReLU $\times2$, Dropout(0.1) 
& $(B,\,512,\,\tfrac{L}{8},\,\tfrac{W}{8})$ \\
\hline
Up2 & $(B,\,512,\,\tfrac{L}{8},\,\tfrac{W}{8})$ 
& Upsample + Double Conv2D 
& ConvTranspose2D $512\to256$; Concat with $(B,\,256,\,\tfrac{L}{4},\,\tfrac{W}{4})$; Conv2D $512\to256$, ReLU $\times2$, Dropout(0.1) 
& $(B,\,256,\,\tfrac{L}{4},\,\tfrac{W}{4})$ \\
\hline
Up3 & $(B,\,256,\,\tfrac{L}{4},\,\tfrac{W}{4})$ 
& Upsample + Double Conv2D 
& ConvTranspose2D $256\to128$; Concat with $(B,\,128,\,\tfrac{L}{2},\,\tfrac{W}{2})$; Conv2D $256\to128$, ReLU $\times2$, Dropout(0.1) 
& $(B,\,128,\,\tfrac{L}{2},\,\tfrac{W}{2})$ \\
\hline
Up4 & $(B,\,128,\,\tfrac{L}{2},\,\tfrac{W}{2})$ 
& Upsample + Double Conv2D 
& ConvTranspose2D $128\to64$; Concat with $(B,\,64,\,L,\,W)$; Conv2D $128\to64$, ReLU $\times2$, Dropout(0.1) 
& $(B,\,64,\,L,\,W)$ \\
\hline
OutConv & $(B,\,64,\,L,\,W)$ 
& Output Conv2D 
& Conv2D $64\to79$, kernel $1\times1$ 
& $(B,\,79,\,L,\,W)$ \\
\hline
\end{tabular}
\end{adjustbox}
\caption{Detailed blockwise architecture of U-Net with a base-channel width of 64.}
\label{tab:A.7}
\end{table}
\end{landscape}
\clearpage

\bibliographystyle{elsarticle-num-names} 
\bibliography{main.bib}

\end{document}